\documentclass[
preprintnumbers,
preprint,
aps,
prd,
superscriptaddress,
nofootinbib,
amsmath
]{revtex4}
\usepackage{bm}
\usepackage[dvips]{color,graphicx}
\usepackage{epsfig}
\usepackage[figuresright]{rotating}
\allowdisplaybreaks[4]
\newcommand{\ud}{\mathrm{d}}
\newcommand{\eq}[1]{Eq.~(\ref{#1})}
\newcommand{\Eqs}[2]{Eqs.~(\ref{#1}) and (\ref{#2})}
\newcommand{\eqs}[1]{Eqs.~(\ref{#1})} 
\newcommand{\dirac}[1]{{\not{\!{#1}}}}

\newcommand{\Tr}{\mathop{\mathrm{Tr}}} 
\newcommand{\tr}{\mathop{\mathrm{tr}}} 
\newcommand{\ceps}{\varepsilon}

\newcommand{\sop}{\widehat{\mathcal{S}}}

\newcommand{\pzhatsq}{{\widehat{p}}_z^2}

\newcommand{\trans}[1]{{#1}^{\prime}}
\newcommand{\intp}{\int[\ud p]}
\newcommand{\sqpsq}{\left( p^2 \right)^{1/2}}
\newcommand{\spfn}[2]{\mathcal{P}^{#1}(\varepsilon,\bm{p},{#2})}
\begin{document}

\preprint{JLAB-THY-08-839}

\title{Spin Structure Functions of $^3$He at Finite $Q^2$}

\author{S. A. Kulagin}
\affiliation{Institute for Nuclear Research,
        Russian Academy of Sciences, 
        117312 Moscow, Russia} 

\author{W. Melnitchouk}
\affiliation{Jefferson Lab,
        12000 Jefferson Avenue,
        Newport News, VA 23606, USA}

\begin{abstract}
Using recently derived relations between spin-dependent nuclear and
nucleon $g_1$ and $g_2$ structure functions at finite $Q^2$, we study
nuclear effects in $^3$He in the nucleon resonance and deep inelastic
regions.
Comparing the finite-$Q^2$ results with the standard convolution
formulas obtained in the large-$Q^2$ limit, we find significant
broadening of the effective nucleon momentum distribution functions,
leading to additional suppression of the nuclear $g_1$ and $g_2$
structure functions around the resonance peaks.
\end{abstract}

\maketitle

\section{Introduction}

The nuclear EMC effect, observed 25 years ago in unpolarized $\mu A$
deep inelastic scattering \cite{EMC}, demonstrated a dramatic change
in the structure function of a nucleon bound in a heavy nucleus
relative to that in the deuteron.
Since then many ideas have been put forward to describe this
modification \cite{EMCrev,KP}, although the exact mechanism(s)
responsible remains controversial.
Recent discussion of the medium modification of nuclear structure
functions has focused on polarization, in the expectation that study
of deep inelastic scattering from polarized nuclei can provide clues
about the nature and origin of the effect.
Indeed, recent calculations \cite{PolEMC} suggest that the
spin-dependent proton $g_1$ structure function undergoes more dramatic
change in the nuclear medium than its unpolarized counterpart, $F_2$.

As a purely practical application, polarized nuclear targets are also
presently the only source of information about the structure of
polarized neutrons, given the absence of free neutron targets.
In particular, polarized $^3$He nuclei are commonly
used as effective neutron targets \cite{He3exp}.
A number of studies of polarized deep inelastic scattering (DIS)
from $^3$He nuclei have been made in recent years
\cite{Wol,Kap,Ciofi93,SS,MPT,KMPW,SSlc,Bissey},
which have attempted to quantify the spin dependence of the
nuclear effects on bound nucleon structure functions.

The standard formalism used to study nuclear DIS at large Bjorken-$x$ 
($x > 0.1$) is the nuclear impulse approximation, in which virtual 
photon--nucleus scattering proceeds at the sub-nuclear level via
virtual photon--nucleon scattering.
In the Bjorken limit (where both the energy transfer $\nu$ and 
four-momentum transfer squared $Q^2$ are large), this formalism
allows nuclear spin structure functions to be expressed as
convolutions of bound nucleon structure functions and spin-dependent
light-cone momentum distributions of nucleon in nuclei, which in this
limit are independent of $Q^2$.

In practice, however, many of the experiments with polarized nuclear
targets are performed at average $Q^2$ values of between $\sim 1$ and
10~GeV$^2$ \cite{He3exp}.
Some analyses of spin-dependent structure functions are made at even
smaller $Q^2$: the generalized Gerasimov-Drell-Hearn (GDH) sum rule,
for instance, interpolates between the deep inelastic region and
the photoproduction limit, where it is given in terms of the nucleon
magnetic moment \cite{GDH}.
It is important, therefore, if one is to accurately describe nuclear
structure function data at current experimental kinematics, and
reliably extract neutron structure information from nuclear targets,
that a framework exist within which one can compute nuclear structure
functions at both high and low $Q^2$.

In a recent paper \cite{DFINQ} we evaluated the effects of finite-$Q^2$
kinematics on the $g_1$ and $g_2$ spin structure functions of the
deuteron, focusing in particular on the resonance region where the
finite-$Q^2$ smearing had significant effects.
The resonance region has received considerable interest recently in
connection with the phenomenon of Bloom-Gilman duality \cite{BG},
which relates structure functions in the resonance and DIS regions
\cite{MEK}.
In particular, it was shown \cite{DFINQ} that at finite $Q^2$ the
simple $Q^2$-independent factorization of the convolution approximation
breaks down, and the effective nucleon momentum distribution functions
acquire an explicit dependence on the scale $Q^2$.
In this work we extend this formalism to the case of inclusive
scattering from the $^3$He nucleus.
Inclusive scattering from polarized $^3$He in the resonance region
was also considered in Ref.~\cite{ScopHe3}, although using a
different formalism.
The differences with our results appear in terms that are higher order 
in the bound nucleon momentum, which may be important in the low-$Q^2$ 
region.

In the following section we review the formalism of the nuclear
impulse approximation, and outline the derivation of the nuclear
hadronic tensor in the approximation of weak nuclear binding.
In Sec.~III we discuss the general properties of the nuclear spectral
function, before turning to the specific case of the $^3$He nucleus.
The complete formulas for the $g_1$ and $g_2$ structure functions of
nuclei in terms of the structure functions of bound nucleons, valid
at finite values of $Q^2$, are presented in Sec.~IV, which will be
particularly useful for studying nuclear effects in the nucleon 
resonance region.
Our equations generalize the Bjorken limit expressions used in earlier
analyses, and are consistent with those for $Q^2 \to \infty$.
In Sec.~IV.B we illustrate the behavior of the spin-dependent
nucleon (light-cone) momentum distributions away from the Bjorken
limit.
Numerical results for the $^3$He structure functions are presented in
Sec.~IV.C, where we compare nuclear effects in both the resonance and
deep inelastic regions using our full calculation with those based on 
various approximations, 
Finally, in Sec.~V we summarize our results and outline future
extensions of this work.

\section{Nuclear Spin-Dependent Structure Functions}
\label{sec:Asf}

This section outlines the derivation of the basic relations between
the nuclear and nucleon hadronic tensors in the nuclear impulse 
approximation.
Starting from a relativistic framework, we systematically apply
the nonrelativistic or weak binding approximation for the nucleon
propagator, which enables the nuclear hadronic tensor to be written
in terms of an off-shell nucleon truncated tensor and a nonrelativistic
nuclear spectral function.
While the general formalism is applicable to nuclei with arbitrary
spin, we will focus here on the specific case of spin-1/2 targets
such as $^3$He.
Further details of the derivation can be found in the Appendices.

\subsection{Hadronic tensor}

To leading order in the electromagnetic coupling constant, the
inclusive differential cross section can be written as a product
of leptonic and hadronic tensors.
The former describes the lepton--photon interaction, while the
latter represents the sum of hadronic matrix elements of the
electromagnetic current $J_\mu$ over all hadronic final states.
From completeness of the final states, the hadronic tensor
$W^A_{\mu\nu}$ of the nucleus can be expressed as the Fourier
transform of the nuclear matrix element of the commutator of two
electromagnetic currents:
\begin{equation}
\label{eq:Wdef}
W^A_{\mu\nu}(P_A,q,S) = \frac{1}{4\pi}\int \ud^4z\ e^{iq\cdot z}\:
\langle P_A,S|\left[ J_\mu(z), J_\nu(0) \right]|P_A,S \rangle\ ,
\end{equation}
where $q$ is the four-momentum transfer, $P_A$ is the momentum of the
target nucleus, and $S$ is the target spin polarization axial-vector,
normalized such that $S^2=-1$ and $P_A \cdot S=0$.
For spin-dependent scattering the relevant component of $W^A_{\mu\nu}$
is antisymmetric in the indices $\mu, \nu$ and can be written in
terms of two structure functions, $g_1^A$ and $g_2^A$:
\begin{eqnarray}
\label{eq:W}
W^A_{\mu\nu}(P_A,q,S)
&=& \frac{M_A}{P_A \cdot q}\
    i\,\epsilon_{\mu\nu\alpha\beta}\, q^\alpha
    \left[ S^\beta\ (g_1^A+g_2^A)
	 - P_A^\beta \frac{S \cdot q}{P_A \cdot q}\, g_2^A \right]\ ,
\end{eqnarray}
where $g_1^A$ and $g_2^A$ are Lorentz-invariant functions of the
Bjorken variable $x_A=Q^2/2P \cdot q$ and the photon virtuality $Q^2$,
with $M_A$ the nuclear mass.
The states are normalized such that
$\langle P_A,S|P'_A,S'\rangle
= 2E_{\bm{P_A}}(2\pi)^3\ \delta(\bm{P_A}-\bm{P_A}')\ \delta_{SS'}$,
in which case the structure functions $g_{1,2}^A$ are dimensionless.
Hermiticity of the electromagnetic current further ensures that the 
structure functions are real.
The nucleon hadronic tensor is similar to that in \eq{eq:W}.

Calculations of nuclear structure functions are usually framed in the
context of the nuclear impulse approximation (IA), in which the virtual
photon scatters incoherently from individual nucleons bound in the
nucleus.
Possible effects which go beyond the impulse approximation include
final state interactions (FSI) between the recoiling nucleus and the
produced hadronic state, meson exchange currents (MEC), and
nuclear shadowing.
Both meson exchange currents and shadowing involve coherent, multiple
scattering effects, which are generally restricted to small values of
$x$, $x \lesssim 0.1$ \cite{Shad}.
Also, since it is scalar, direct scattering from a pion in the nucleus
does not contribute to spin-dependent structure functions
(but can contribute of course to polarization asymmetries).

The effects of FSI and MEC have been considered for quasi-elastic (QE)
scattering from $^3$He within a nonrelativistic Faddeev approach in 
Ref.~\cite{Golak}.
Comparison with recent data from Jefferson Lab \cite{Slifer} found FSI
effects to be important at low $Q^2$ ($Q^2 \sim 0.05-0.2$~GeV$^2$),
and gradually decreasing as $Q^2$ increases, bringing the IA
calculations closer to the data.
In this context we also mention the results of Ref.\cite{HLM80},
where in a nonrelativistic Green's function approach analyticity and
unitarity requirements were used to argue the FSI effects to cancel
in energy-integrated inclusive spin-averaged cross sections.
Similar arguments also lead to partial cancellation of FSI effects
in inclusive inelastic cross sections.
Note that the corresponding FSI effects are significantly stronger
for exclusive channels, in which the nucleon is detected in the final
state \cite{HLM80}.
From the existing approaches it is not clear, however, whether FSI 
effects can be neglected in spin-dependent inclusive cross sections
in the resonance region and at higher energies.
Computation of the FSI effects here will require extension of the 
formalism to include relativistic effects and couplings between 
different open channels --- a problem which remains an important 
challenge.

Within the impulse approximation framework the nuclear hadronic tensor
can be written as:
\begin{equation}
\label{eq:W:kern}
W^A_{\mu\nu}(P_A,q,S)
= \sum_{\tau=p,n} \int [\ud p]\, 
  \Tr\left[
	{\cal A}^\tau(p,P_A,S)\
	\widehat{\cal W}_{\mu\nu}^\tau(p,q)
     \right]\ ,
\end{equation}
where the integration is performed over the bound nucleon four-momentum
$p$, and we use the shorthand notation $[\ud p] \equiv d^4p/(2\pi)^4$.
Here, and in the following, the index $\tau$ labels the nucleon isospin 
state and a sum is taken over protons ($\tau=p$) and neutrons ($\tau=n$).
The truncated or off-shell nucleon tensor
$\widehat{\cal W}_{\mu\nu}^\tau(p,q)$ describes the inclusive
scattering of the virtual photon from an off-mass-shell nucleon,
and is a matrix in Dirac space (see Appendix~\ref{app:Noff}).
The Dirac matrix ${\cal A}^\tau(p,P_A,S)$ is the imaginary part of the
nucleon propagator in the nucleus $A$ with momentum $P_A$ and spin $S$:
\begin{equation}
\label{eq:A}
{\cal A}_{\alpha\beta}^\tau(p,P_A,S)
= \int \ud t\ \ud^3\bm{r}\ e^{i (p_0 t - \bm{p}\cdot\bm{r})}\
  \langle P_A,S|\
        \overline{\Psi}_\beta^\tau(t,\bm{r})\,\Psi_\alpha^\tau(0)\
  | P_A,S \rangle\ ,
\end{equation}
where $\Psi_\alpha^\tau(t,\bm{r})$ is the (relativistic) nucleon
field operator, and $\alpha, \beta$ are Dirac spinor indices.
The trace ``Tr'' in \eq{eq:W:kern} is taken in the nucleon Dirac space.

\subsection{Weak binding approximation (WBA)}
\label{ssec:WBA}

The expression for the nuclear tensor in Eq.~(\ref{eq:W:kern})
is covariant and can be evaluated in any frame.
It will be convenient, however, to work in the target rest frame,
in which the target momentum is $P_A = (M_A, \bm{0})$ and the
spin vector $S = (0, \bm{S})$, and the momentum transfer to the
target defines the $z$-axis, $q=(q_0, \bm{0}_\perp, -|\bm{q}|)$.
If the nucleus can be approximated as a nonrelativistic system
of weakly bound nucleons with four-momentum
$p \equiv (M+\varepsilon, \bm{p})$, where $M$ is the nucleon mass,
then the nuclear hadronic tensor in \eq{eq:W:kern} simplifies
considerably.
This necessarily involves neglecting antinucleon degrees of freedom,
and corresponds to bound nucleons in the nucleus having small momentum 
and energy, $|\bm{p}|, |\varepsilon| \ll M$.
We refer to this as the ``weak binding approximation'' (WBA).

To proceed, we perform a nonrelativistic reduction of all
Lorentz--Dirac structures in the nucleon hadronic tensor.
This can be done by relating the relativistic four-component nucleon
field $\Psi^\tau$ to the nonrelativistic two-component operator 
$\psi^\tau$:
\begin{equation}
\label{eq:N-psi}
\Psi^\tau(\bm{p},t)
= e^{-iMt}
\left(
  \begin{array}{r}
    Z\,\psi^\tau(\bm{p},t) \\
    \frac{\displaystyle
    \bm{\sigma}\cdot\bm{p}}
    {\displaystyle 2M}\,\psi^\tau(\bm{p},t)
  \end{array}
\right),
\end{equation}
which is valid to order $\bm{p}^2/M^2$.
The validity of \eq{eq:N-psi} relies on the absence of strong fields
in the nucleus comparable to the nucleon mass (see {\em e.g.} the
discussion in the Appendix of Ref.~\cite{KMPW}).
The nucleon operators in \eq{eq:N-psi} are taken in a mixed $(\bm{p},t)$
representation,
$\psi^\tau(\bm{p},t) = \int d^3\bm{r} \exp(-i \bm{p} \cdot \bm{r})
			\psi^\tau(\bm{r},t)$.
The renormalization operator $Z=1-\bm{p}^2/8M^2$ ensures that the
charge (baryon number) is not renormalized when going to the
nonrelativistic limit:
\begin{equation}
\label{eq:norm}
\int \ud^3\bm{p}\,\overline{\Psi}^\tau(\bm{p},0) \gamma_0
		  \Psi^\tau(\bm{p},0)
= \int \ud^3\bm{p}\,\psi^{\dagger \tau}(\bm{p},0)
		    \psi^\tau(\bm{p},0)\ .
\end{equation}
One can then define the nuclear spin-dependent spectral function 
${\cal P}^\tau$ of a nucleus in terms of the correlator of the 
nonrelativistic fields $\psi^\tau$ as:
\begin{equation}
\label{eq:spfn}
\mathcal{P}^{\tau}_{\sigma\sigma'}(\varepsilon,\bm{p},\bm{S})
= \int \ud t\,
  e^{-i\,\varepsilon t}
  \langle A, \bm{S} |
	  \psi^{\dagger \tau}_{\sigma'}(\bm{p},t)
	  \psi^\tau_\sigma(\bm{p},0)
  | A, \bm{S}  \rangle\ ,
\end{equation}
where the expectation value is taken with respect to the nuclear ground 
state $|A, \bm{S}\rangle$, normalized to unity, with polarization 
$\bm{S}$.
The operator
${\psi^{\dagger}}^\tau_{\sigma}(\bm{p},t)\
 (\psi^\tau_{\sigma}(\bm{p},t))$
creates (annihilates) a nonrelativistic nucleon with isospin $\tau$,
momentum $\bm{p}$ and polarization $\sigma$, at time $t$.

Substituting Eq.~(\ref{eq:N-psi}) into Eq.~(\ref{eq:W:kern}), the
four-dimensional Dirac spinor matrices reduce to nonrelativistic
two-dimensional spinors (see Appendix~\ref{app:WBA}):
\begin{align}
\label{eq:Tr:NR}
\frac1{M_A}\,
\Tr\left[
     \widehat\mathcal{W}_{\mu\nu}^\tau(p,q)\
     {\cal A}^\tau(p,S)\,
   \right]
= \frac1{M+\varepsilon}\,
\tr\left[
     \widehat{w}_{\mu\nu}^\tau(p,q)\
     \spfn{\tau}{\bm{S}}\, 
   \right]\ ,
\end{align}
where ``tr'' is now a two-dimensional trace in Pauli matrix space.
The off-shell truncated nucleon tensor
$\widehat{w}_{\mu\nu}^\tau(p,q)$ 
in the vicinity of the mass-shell can be written in a similar way
to the relativistic nucleon tensor in Eq.~(\ref{eq:W}):
\begin{equation}
\label{eq:W:NR}
\widehat{w}_{\mu\nu}^\tau(p,q)
= \frac{M}{p\cdot q} i
  \epsilon_{\mu\nu\alpha\beta}\, q^\alpha
  \left[ \sop^\beta\ (g_1^\tau + g_2^\tau)
	- p^\beta\ \frac{\sop\cdot q}{p\cdot q}\ g_2^\tau
  \right]\ ,
\end{equation}
where $g_{1,2}^\tau$ are the structure functions of the proton or
neutron with four-momentum $p$.
Note that these functions generally depend on 3 variables:
$Q^2$, $x'=Q^2/2p\cdot q$ and $p^2$.
The operator $\sop$ has a structure similar to that of the spin
four-vector $(0,\bm{\sigma})$ boosted to a frame in which the
nucleon has the (nonrelativistic) momentum $\bm{p}$:
\begin{equation}
\label{eq:S:NR}
\sop = \left(\frac{\bm{\sigma}\cdot\bm{p}}{M},\,
\bm{\sigma}+\frac{\bm{p}\,(\bm{\sigma}\cdot\bm{p})}{2M^2}\right)\ .
\end{equation}
Combining Eqs.~(\ref{eq:Tr:NR}) and (\ref{eq:W:kern}) leads then
to the relation between the nuclear and nucleon tensors:
\begin{equation}
\label{eq:W:kern:NR}
\frac{1}{M_A} W_{\mu\nu}^A(P_A,q,S)
= \sum_{\tau} \int \frac{[\ud p]}{M+\varepsilon}
  \tr\left[
       \widehat{w}_{\mu\nu}^\tau(p,q)\ 
       \spfn{\tau}{\bm{S}}
     \right]\ .
\end{equation}
In the derivation of Eq.~(\ref{eq:Tr:NR}) all terms have been kept to
order $\bm{p}^2/M^2$ and $\varepsilon/M$, with higher order terms
neglected.
On the other hand, no assumption has been made about the scale $Q^2$,
so that Eq.~(\ref{eq:Tr:NR}) holds for arbitrary values of $Q^2$,
so long as the nuclear impulse approximation is valid.
The relation between the nuclear and nucleon structure functions,
which follow from \eq{eq:W:kern:NR}, will be discussed in
Sec.~\ref{sec:He3} below.

\section{Nuclear spectral function}
\label{sec:spfn}

In this section we present a detailed discussion of the nuclear
spectral function.
After outlining its general properties for an arbitrary nucleus,
we then focus on the specific case of $^3$He.

\subsection{General properties}
\label{ssec:gen}

The nuclear spectral function ${\cal P}^\tau$ in Eq.~(\ref{eq:spfn})
is a matrix in the nucleon spin space.
The general spin structure of the spectral function can be obtained
by expanding ${\cal P}^\tau$ in terms of the Pauli spin matrices and
applying constraints from parity and time-reversal invariance
\cite{SS}:
\begin{equation}
\label{eq:spfn:2}
\spfn{\tau}{\bm{S}} = \frac12 
\left(
   f_0^\tau\ {I}
 + f_1^\tau\,\bm{\sigma}\cdot\bm{S}
 + f_2^\tau\, T_{ij}\, S_i\, \sigma_j
\right)\ ,
\end{equation}
where $I$ is the $2 \times 2$ identity matrix,
$T_{ij} = \widehat p_i\ \widehat p_j - \tfrac13\ \delta_{ij}$
is a traceless symmetric tensor, with $\widehat p_i=p_i/|\bm{p}|$
the $i$-th spatial component of the momentum, and a sum over
repeated indices is implied.

From the hermiticity of the spectral function, the coefficients in 
\eq{eq:spfn:2} are real functions of the energy $\ceps$ and momentum 
$\bm{p}$, $f_m^\tau = f_m^\tau(\ceps,\bm{p})$, with $m=0,1,2$.
The function $f_0^\tau$ describes the spin-averaged spectral function,
while $f_1^\tau$ and $f_2^\tau$ characterize the nucleon spin
distributions in the nucleus.
The function $f_0^p$ $(f_0^n)$ is normalized to the number of
protons (neutrons) in the nucleus:
\begin{equation}
\int [\ud p]\ \tr\left[\spfn{p(n)}{S} \right]
= \int [\ud p]\, f_0^{p(n)} = Z\ (A-Z)\ ,
\label{eq:f0norm}
\end{equation}
which follows from Eqs.~(\ref{eq:norm}) and (\ref{eq:spfn}).
The usual nucleon ``momentum distribution'' in the nucleus is given
by the integral of $f_0^\tau$ over $\ceps$:
\begin{equation}
n^\tau(\bm{p})
= \int \frac{d\ceps}{2\pi}\ f_0^\tau(\ceps,\bm{p})\ .
\end{equation}
The integrated function $f_1^\tau$ determines the average nucleon
polarization in the nucleus:
\begin{equation}
\label{eq:f1norm}
\left\langle \sigma_z \right\rangle^\tau
= \intp\, f_1^\tau\ ,
\end{equation}
while $f_2^\tau$ is related to the tensor polarization:
\begin{equation}
\label{eq:f2norm}
\left\langle T_{zi}\ \sigma_i \right\rangle^\tau
= \frac29 \intp\, f_2^\tau\ ,
\end{equation}
where we define the nuclear spin vector $\bm{S}$ to lie along the 
$z$-axis.
Using the expansion of ${\cal P}^\tau$ in \eq{eq:spfn:2}, the traces on 
the right-hand-side of \eq{eq:W:kern:NR} can be written in terms of the
functions $f_{1,2}^\tau$:
\begin{subequations}
\label{eq:Tr:spfn}
\begin{eqnarray}
\tr\left[ \spfn{\tau}{\bm{S}}\, \sop_0 \right]
&=& \left( f_1^\tau + \frac23 f_2^\tau \right)
    \bm{S} \cdot \bm{v}\ ,				\\
\tr\left[ \spfn{\tau}{\bm{S}}\, \sop_i \right]
&=& \left[ f_1^\tau + \frac16 v^2 \left( f_1^\tau + \frac23 f_2^\tau
				                  \right)
    \right] S_i\					\nonumber\\
&+& \left[ f_2^\tau + \frac12 v^2 \left( f_1^\tau + \frac23 f_2^\tau
				                  \right)
    \right] T_{ij} S_j\ ,
\end{eqnarray}
\end{subequations}
where $\bm{v} \equiv \bm{p}/M$ is the nucleon velocity.

Inserting a complete set of intermediate states in \eq{eq:spfn} and
calculating the transition matrix elements between the ground and
intermediate states allows the spectral function to be written as:
\begin{equation}
\label{eq:spfn4}
\mathcal{P}^\tau_{\sigma\sigma'}(\varepsilon,\bm{p})
= \sum_f
  \psi^\tau_{f,\sigma}(\bm{p}) \psi^{*\ \tau}_{f,\sigma'}(\bm{p})\, \
  2\pi\ \delta \left[ \varepsilon + M + E_{A-1}^f(\bm{p}) - M_A \right]\ ,
\end{equation}
where the function
$\psi_{f,\sigma}^\tau(\bm{p})
= \left\langle (A-1)_f,-\bm{p}|\psi_\sigma^\tau(0)|A \right\rangle$
gives the probability amplitude to find a nucleon with isospin $\tau$ 
and polarization $\sigma$ in the nuclear ground state and the 
remaining $A-1$ nucleons in a state with total momentum $-\bm{p}$,
with the subscript $f$ labeling all other quantum numbers.
The energy of the residual system, including the kinetic energy, is 
denoted by $E_{A-1}^f(\bm{p}) = M_{A-1}^f + \bm{p}^2/2M_{A-1}^f$,
where $M_{A-1}^f$ is the mass of remaining $(A-1)_f$ nuclear system.

Note that \eq{eq:spfn4} is written in the target rest frame and
defines the nuclear spectral function as a function of nucleon
energy $\ceps=p_0-M$ and momentum $\bm p$.
However, in practice the spectral function is usually considered
as a function of the nucleon separation energy $E$.
The spectral function in this case, denoted by $P(E,\bm p)$, is
given by \eq{eq:spfn4} with the energy conserving $\delta$-function
replaced by $\delta(E-E^f)$, where
\begin{equation}\label{eq:Ef}
E^f = M + M_{A-1}^f - M_A
\end{equation}
is the energy needed to separate a nucleon from the nucleus $A$,
leaving the residual nuclear system in a state $(A-1)_f$.
For a given (positive) separation energy $E$, the relation between
the two spectral functions $\mathcal{P}(\varepsilon,\bm{p})$ and
$P(E,\bm p)$ can be found by defining
\begin{equation}
\label{eq:Edef}
\ceps(E,\bm p) = -E - \frac{\bm p^2}{2(M_{A}-M+E)}\ ,
\end{equation}
as suggested by \eq{eq:Ef} and the argument of the energy
$\delta$-function in \eq{eq:spfn4}.
From \Eqs{eq:spfn4}{eq:Edef} one then finds the relation
\begin{equation}\label{eq:spfn5}
P(E,\bm p) = \left| \frac{\partial\ceps(E,\bm p)}{\partial E}\right|
			  \mathcal{P}(\ceps(E,\bm p),\bm{p})\ ,
\end{equation}
where the derivative factor ensures that the energy integrals of the
two spectral functions are equal, {\em i.e.},
$\int \ud E\, P(E,\bm p) = \int \ud \ceps\, \mathcal P(\ceps,\bm p)$,
and the normalizations in Eqs.~(\ref{eq:f0norm})--(\ref{eq:f2norm})
remain valid.
In the following we shall take the functions $f_m^\tau$ to be
functions of $E$ and $\bm p$.

For the deuterium nucleus, discussed in Ref.~\cite{DFINQ}, the
intermediate states are exhausted by a single proton or neutron.
In this case the spectral function is expressed entirely in terms
of the deuteron wave function.
For the three-nucleon system, however, the calculation of the
spectral function is rather more complicated, as we discuss next.

\subsection{$^3$He spectral function}
\label{ssec:wfns}

For a $^3$He nucleus, the proton spectral function has two contributions:
from the bound $(pn)$ intermediate state corresponding to a deuteron,
where the separation energy is $E = \ceps_d - \ceps_{^3{\rm He}}$, with
$\ceps_d = -2.22$~MeV and $\ceps_{^3{\rm He}} = -7.72$~MeV the deuterium
and $^3$He binding energies; and from the $(pn)$ continuum scattering 
states, with energy $E$.
The proton functions $f_m^p$ can be written as
\begin{eqnarray}
\label{eq:fpHe3}
f_m^p(E,\bm{p})
&=& f_m^{p (d)}(\bm{p})
    \delta\left( E + \ceps_{^3{\rm He}} - \ceps_d \right)\
 +\ f_m^{p ({\rm cont})}(E,\bm{p})\ .
\end{eqnarray}
The neutron spectral function, on the other hand, has only the
$(pp)$ continuum contribution:
\begin{eqnarray}
\label{eq:fnHe3}
f_m^n(E,\bm{p})
&=& f_m^{n ({\rm cont})}(E,\bm{p})\ .
\end{eqnarray}
From these spectral functions one can compute the effective polarization
and tensor polarization of the proton and neutron in the $^3$He nucleus
using Eqs.~(\ref{eq:f1norm}) and (\ref{eq:f2norm}).

More familiar notation expresses the polarizations in terms of
probabilities to find the proton and neutron in various orbital states.
For $^3$He these correspond to the dominant space-symmetric
(spin-isospin antisymmetric) $S$-state ($P_S$);
a small (1--2\%) admixture of the $L=0$ mixed-symmetric $S'$ state
($P_{S'}$), which reflects spin-isospin correlations in the nuclear
force;
and the $L=2$ $D$-state ($P_D$), generated by the tensor force,
in which the nucleon spins are aligned antiparallel to the nuclear
spin projection, with probability $\sim 10\%$ \cite{Friar}.
The probability of the $L=1$ $P$-state is very small ($<1\,$\%)
and is not considered here.
The effective proton and neutron polarizations in $^3$He can then
be written as \cite{Friar}:
\begin{subequations}
\label{eq:np-pol}
\begin{eqnarray}
\langle \sigma_z \rangle^p
&=& -\frac23\left(P_D-P_{S'}\right)\ , \\
\langle \sigma_z \rangle^n
&=& P_S-\frac13\left(P_D - P_{S'}\right)\ ,
\end{eqnarray}
\end{subequations}
where $\langle \sigma_z \rangle^p$ is defined as the {\em total}
proton polarization ({\em i.e.}, the sum of the two protons'
polarizations).

In the present analysis we use the spectral functions from
Refs.~\cite{KPSV,Salme} (denoted by ``KPSV'') and Ref.~\cite{SS}
(denoted by ``SS''), which provide a representative sample of models
of the $^3$He nucleus.
The KPSV spectral function \cite{KPSV} is obtained using a $^3$He 
wave function calculated in a variational method with a 
pair-correlated hyperspherical harmonic basis \cite{KVR}, for various 
$NN$ potentials, including a three-body force and a Coulomb interaction 
between the two protons.
In our numerical calculations we use the KPSV spectral function with
the more recent AV18 $NN$ potential and the $NNN$ Urbana IX interaction 
\cite{Salme}.
The SS spectral function \cite{SS}, on the other hand, is obtained by
solving the Faddeev equation with the Paris $NN$ potential \cite{SSG}
for the ground state wave function, and constructing its projection
onto the deuteron and two-body continuum states.

The neutron polarization from the KPSV (SS) spectral function is
$\langle \sigma_z \rangle^n = 0.86$ (0.89).
The deuteron pole contribution to the proton polarization is large
and negative,
	$\langle \sigma_z \rangle^{p(d)} = -0.445\ (-0.453)$
for the KPSV (SS) model, but is canceled by the positive continuum
contribution,
	$\langle \sigma_z \rangle^{p(\rm cont)} = 0.386\ (0.409)$,
leaving a very small negative total proton polarization,
	$\langle \sigma_z \rangle^p = -0.059\ (-0.044)$
for the KPSV (SS) spectral function.
Note that the deuteron pole contribution to the spin-averaged
spectral function is rather large, providing some 67\% (68\%)
of its normalization in the KPSV (SS) model.
The corresponding tensor polarizations of the neutron for the two
models are
$\langle T_{zi} \sigma_i \rangle^n = 0.045$ \cite{Salme} and
0.038 \cite{SS}, respectively.
The proton tensor polarizations,
$\langle T_{zi} \sigma_i \rangle^p = -0.226$ \cite{Salme} and
$-0.235$ \cite{SS},
arise mainly from the deuteron pole, with the continuum contributions
being $< 2\%$ and $< 10\%$ for the KPSV and SS spectral functions,
respectively.

\section{Polarized $^3$He structure functions}
\label{sec:He3}

In this section we apply the formalism developed in Sec.~\ref{sec:Asf}
together with the spectral functions described in Sec.~\ref{sec:spfn}
to compute the $g_1$ and $g_2$ structure functions of $^3$He.
We present the complete finite-$Q^2$ formulas for the nuclear structure
functions, and also formulate the results in terms of the familiar
light-cone convolution formulas.
We conclude by performing a detailed analysis of the nuclear effects on 
the nucleon structure functions in both the resonance and DIS regions.

\subsection{Finite-$Q^2$ convolution}
\label{ssec:master}

Using the nuclear hadronic tensor in \eq{eq:W:kern:NR} we can write
the nuclear structure functions $g_1^A$ and $g_2^A$ as convolutions
of the (off-shell) nucleon spin structure functions and the nuclear 
spectral function.
The nuclear and nucleon structure functions can be related by
considering  the helicity structure functions in
\Eqs{eq:W:h:1}{eq:W:h:01} of  Appendix~\ref{app:hel}, and projecting
onto appropriate helicity states.
The final result for the nuclear $g_1^A$ and $g_2^A$ structure 
functions at finite $Q^2$ can be summarized as:
\begin{equation}
\label{eq:master}
xg_a^A(x,Q^2)
= \int [\ud p]\, D_{ab}^\tau(\ceps,\bm{p},\gamma)\, 
                 x' g_b^\tau(x',Q^2,p^2)\ ,
\end{equation}
where $x' = Q^2/2p \cdot q = x/[1+(\ceps+\gamma p_z)/M]$
is the Bjorken variable for the bound nucleon,
$p^2=(M+\ceps)^2-\bm{p}^2$ is the off-shell nucleon virtuality,
and $a,b=1,2$ (in order to simplify notation, summation over
repeated indices $b$ and $\tau$ is assumed).
The energy-momentum distributions $D_{ab}$ are given by:
\begin{subequations}
\label{eq:Dab}
\begin{align}
D_{11} &=
     f_1 
   + \frac{3-\gamma^2}{6\gamma^2}\left(3\pzhatsq - 1\right) f_2
  + \frac{v\widehat p_z}{\gamma}\left(f_1 + \tfrac23 f_2\right)
 + v^2\frac{(3-\gamma^2)\pzhatsq - 1-\gamma^2}{12\gamma^2}(3f_1-f_2), 
\label{eq:D11}\\
D_{12} &= 
     (\gamma^2-1)
\left[
         - \frac{3\pzhatsq - 1}{2\gamma^2}\, f_2
      + \frac{v\widehat p_z}{\gamma}
        \left(f_1+\left(\tfrac32\pzhatsq - \tfrac56\right)\,f_2\right)
\right.
\notag
\\ &
\left.
    {} - v^2 \left(
  \frac{1+\pzhatsq (4\gamma^2-3)}{4\gamma^2}\,f_1
+\frac{5+18\widehat{p}_z^4\gamma^2 - 5\pzhatsq (3+2\gamma^2)}{12\gamma^2}\,f_2
            \right) 
\right],
\label{eq:D12}\\
D_{21} &=
     -\frac{3\pzhatsq - 1}{2\gamma^2} f_2
    -\frac{v\widehat{p}_z}{\gamma}\left(f_1 + \tfrac23 f_2 \right)
  -v^2 \frac{3\pzhatsq - 1}{12\gamma^2}\left(3f_1-f_2 \right),
\label{eq:D21}\\
D_{22} &=
     f_1 + \frac{2\gamma^2-3}{6\gamma^2}\left(3\pzhatsq - 1\right)f_2
   +\frac{v\widehat{p}_z}{\gamma}
     \left[
     (1-\gamma^2) f_1
	+ \left( -\tfrac56 + \tfrac13 \gamma^2
		 + \pzhatsq (\tfrac32 - \gamma^2)
	  \right) f_2
     \right]
\notag \\
 & {}
+ v^2
\left[
   \frac{\pzhatsq (3-6\gamma^2+4\gamma^4) - 1 - 2\gamma^2}
	{4\gamma^2}\, f_1 
 + \frac{5 - 2\gamma^2 (1 + 3\pzhatsq) + 4\pzhatsq\gamma^4}
	{12\gamma^2} (3\pzhatsq - 1) f_2
\right] ,
\label{eq:D22}
\end{align}
\end{subequations}
where $v = |\bm{v}| = |\bm{p}|/M$,
$\gamma = |\bm{q}|/q_0 = \sqrt{1+(2Mx)^2/Q^2}$ is the ``velocity''
of the virtual photon, and for clarity the isospin indices have been 
suppressed.
The expressions in \eqs{eq:Dab} are derived in the weak binding
approximation, making use of the fact that characteristic values
of the nucleon velocity $v$ are small, which allows the kinematical
factors to be expanded up to order $v^2$.

Note that while in the Bjorken limit the nucleon distributions
$D_{ab}$ are independent of $Q^2$, at finite $Q^2$ the scale
dependence enters explicitly through the parameter $\gamma$.
Furthermore, nuclear effects lead to nonzero off-diagonal
distributions $D_{12}$ and $D_{21}$, giving rise to the mixing
of different spin structure functions in the convolution integral
(\ref{eq:master}).
In the limit of high $Q^2$ the parameter $\gamma \to 1$ and the
distributions (\ref{eq:Dab}) simplify considerably.
In particular, in this limit the function $D_{12} \to 0$,
and the convolution formula for $g_1^A$ becomes diagonal.
On the other hand, the mixing effect for $g_2^A$ persists
even in the $\gamma \to 1$ limit \cite{Ciofi93,KMPW}.
The nuclear $g_1^A$ and $g_2^A$ structure functions can then
be computed from Eqs.~(\ref{eq:Dab}) at any value of $Q^2$.

The relation between the nuclear and nucleon spin structure functions
for polarized $^3$He has also been discussed in several previous
studies \cite{SS,Ciofi93,SSlc,ScopHe3}. 
The implementation of the impulse approximation in Refs.~\cite{SS,SSlc}
is different from the present approach, leading to different
distribution functions in the nuclear convolution. 
In particular, in Ref.~\cite{SS,SSlc} the struck nucleon was assumed
to be on-mass-shell, while in our approach the nucleon is by
definition off its mass-shell.
This leads to different definitions of the Bjorken variable of the 
struck nucleon $x'$, resulting in significant differences in the 
numerical analysis.
The convolution equation \eq{eq:master} and the corresponding
distribution in \eq{eq:Dab} for $g_1$ are similar to those in
Ref.~\cite{Ciofi93,ScopHe3} to leading order in $v$ and in the
Bjorken limit $\gamma \to 1$.
However, the higher order terms in $v$ and $\gamma^2-1$ are different.

\subsection{Light-cone distributions}
\label{ssec:Dy}

At large $Q^2$ deep inelastic nuclear structure functions are usually
written as simple, one-dimensional convolutions of nucleon structure
functions and nucleon light-cone momentum distribution functions in
the nucleus.
We can generalize this convolution to finite $Q^2$ by using 
Eq.~(\ref{eq:master}) and defining effective light-cone momentum
distributions $f_{ab}^\tau$ as integrals of the functions
$D_{ab}^\tau$ in Eq.~(\ref{eq:Dab}):
\begin{eqnarray}
\label{eq:Dy}
f_{ab}^\tau(y,\gamma)
&=& \int [\ud p]\ D_{ab}^\tau(\ceps,\bm{p},\gamma)\
    \delta \left( y-1-\frac{\ceps +\gamma p_z}{M} \right)\ ,
\end{eqnarray}
where the variable
$y = (p_0 + \gamma p_z)/M
   = 1 + (\varepsilon+\gamma p_z)/M = x/x'$
in the Bjorken ($\gamma\to 1$) limit is the light-cone fraction
of the nucleus carried by the interacting nucleon. 
The nuclear structure functions $g_1^A$ and $g_2^A$ can then be
written as \cite{DFINQ}
\begin{eqnarray}
\label{eq:convLC}
x g_a^A(x,Q^2)
&=& \int_x^{M_A/M} \ud y\ f_{ab}^\tau(y,\gamma)\
    x' g_b^\tau\left(x',Q^2\right)\ .
\end{eqnarray}
For inelastic scattering the lower limit of the $y$-integration is
$y_{\rm min} = x/x_{\rm th}$, where 
$x_{\rm th}=[1+(2M m_\pi + m_\pi^2)/Q^2]^{-1}$ corresponds to the
inelastic threshold for which the invariant mass of the final
state $W \geq M+m_\pi$.
In Eq.~(\ref{eq:convLC}) we have also suppressed the dependence of
the nucleon structure functions on the nucleon virtuality $p^2$.
We justify this in this analysis since the aim here is to study the
role of the finite-$Q^2$ smearing on structure functions, the effects
of which should be largely independent of details of the nucleon
structure function input.

\begin{sidewaysfigure}[p]
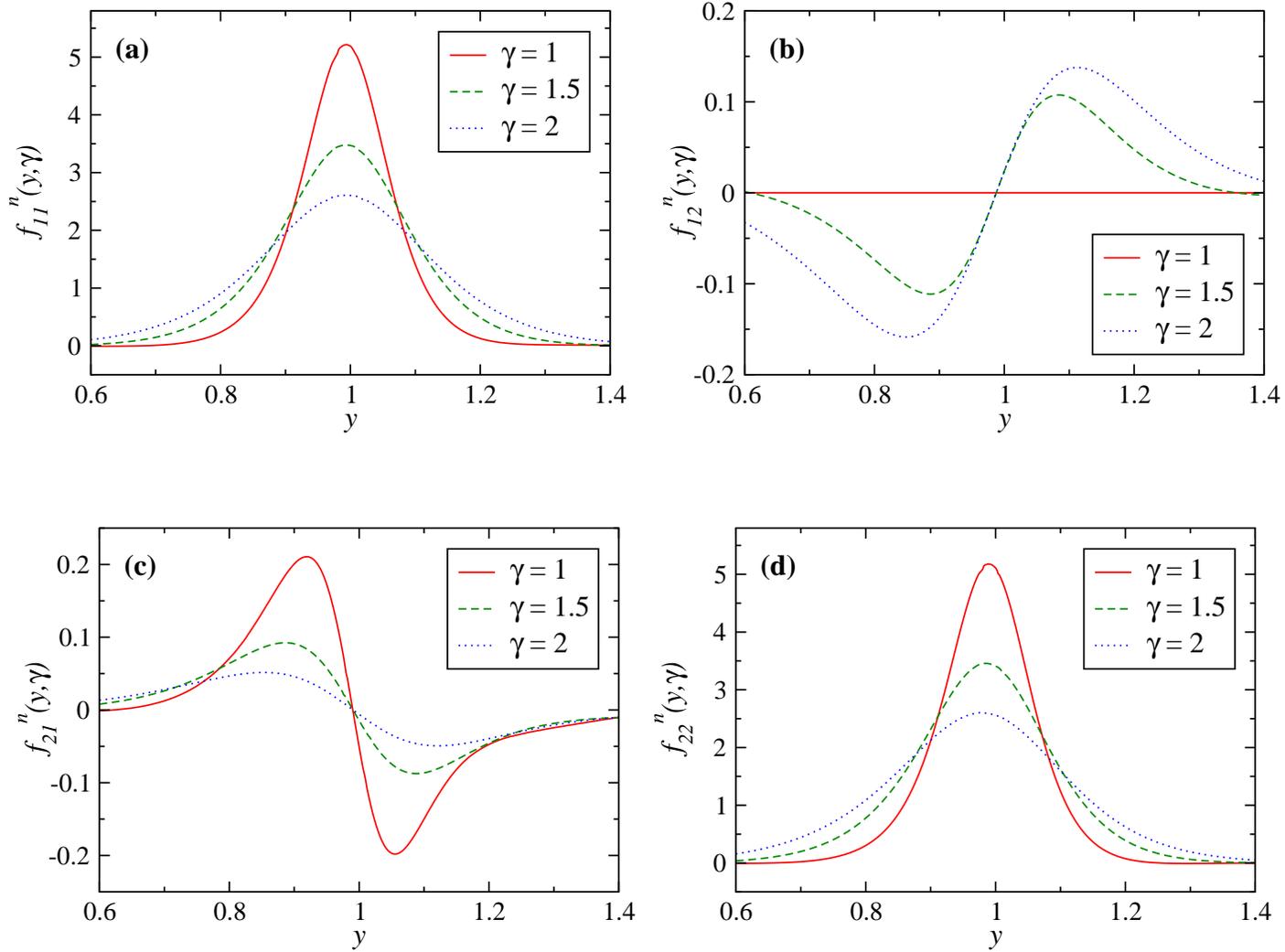

\begin{center}
\epsfig{file=fig1a.eps,width=9cm}\ \ \ \ %
\epsfig{file=fig1b.eps,width=9cm}\vspace*{1.3cm}
\epsfig{file=fig1c.eps,width=9cm}\ \ %
\epsfig{file=fig1d.eps,width=9cm}
\caption{(Color online)
	Neutron light-cone momentum distribution functions
	$f_{ab}^n(y,\gamma)$ in $^3$He,
	for $\gamma=1$ (Bjorken limit), 1.5 and 2.}
\label{fig:1}
\end{center}
\end{sidewaysfigure}

\begin{sidewaysfigure}[p]
\begin{center}
\epsfig{file=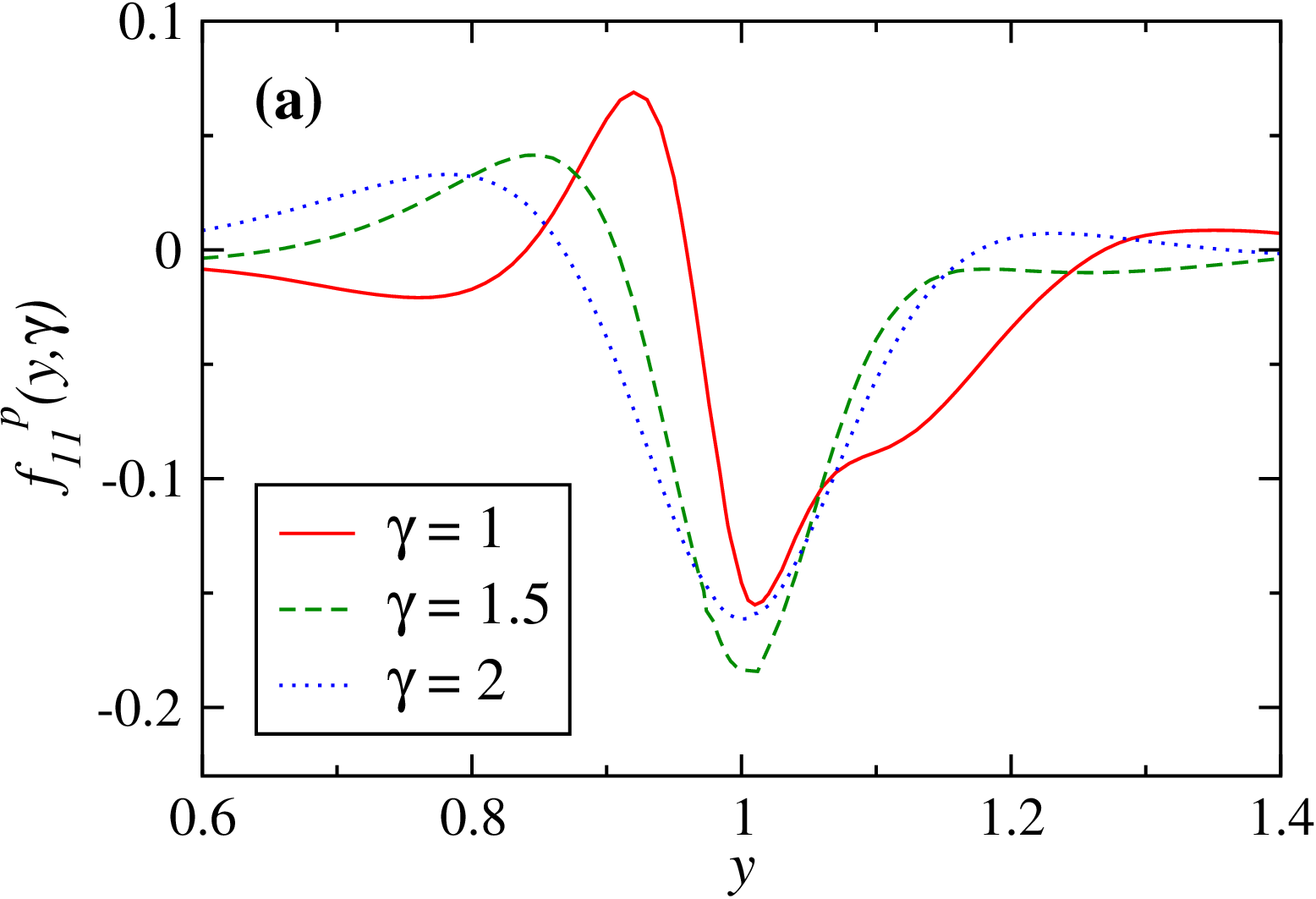,width=9cm}\ \ \ \ %
\epsfig{file=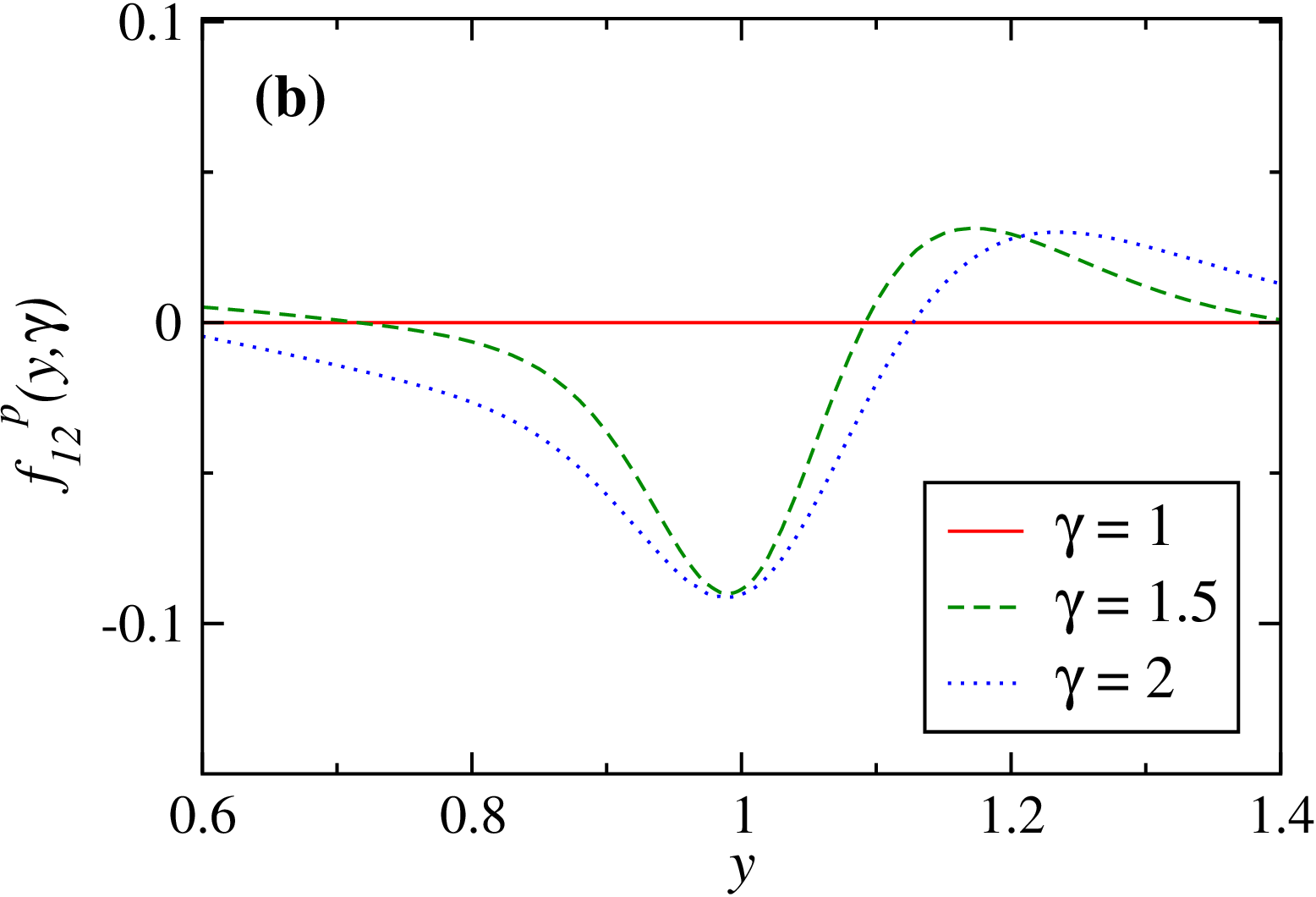,width=9cm}\vspace*{1.3cm}
\epsfig{file=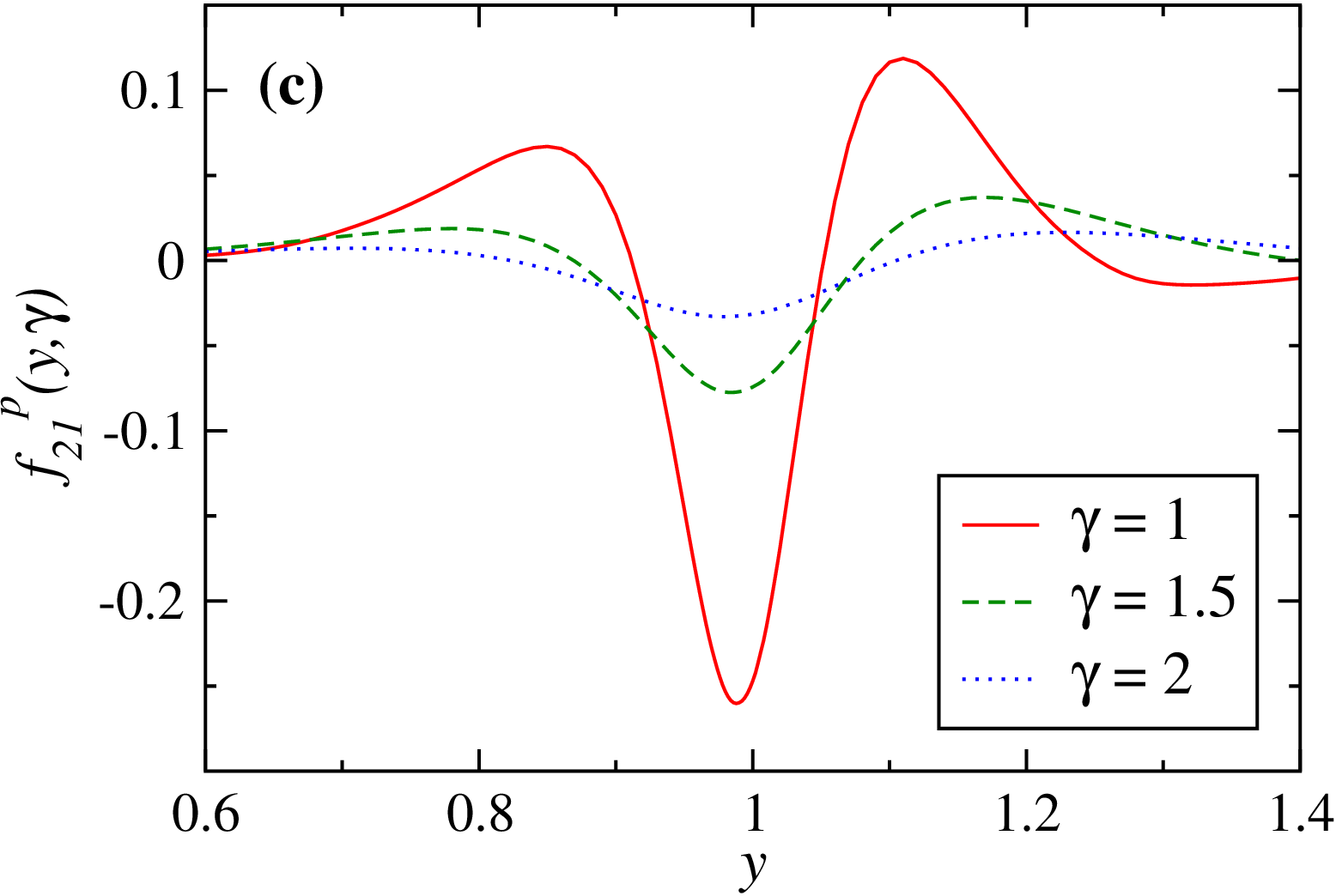,width=9cm}\ \ %
\epsfig{file=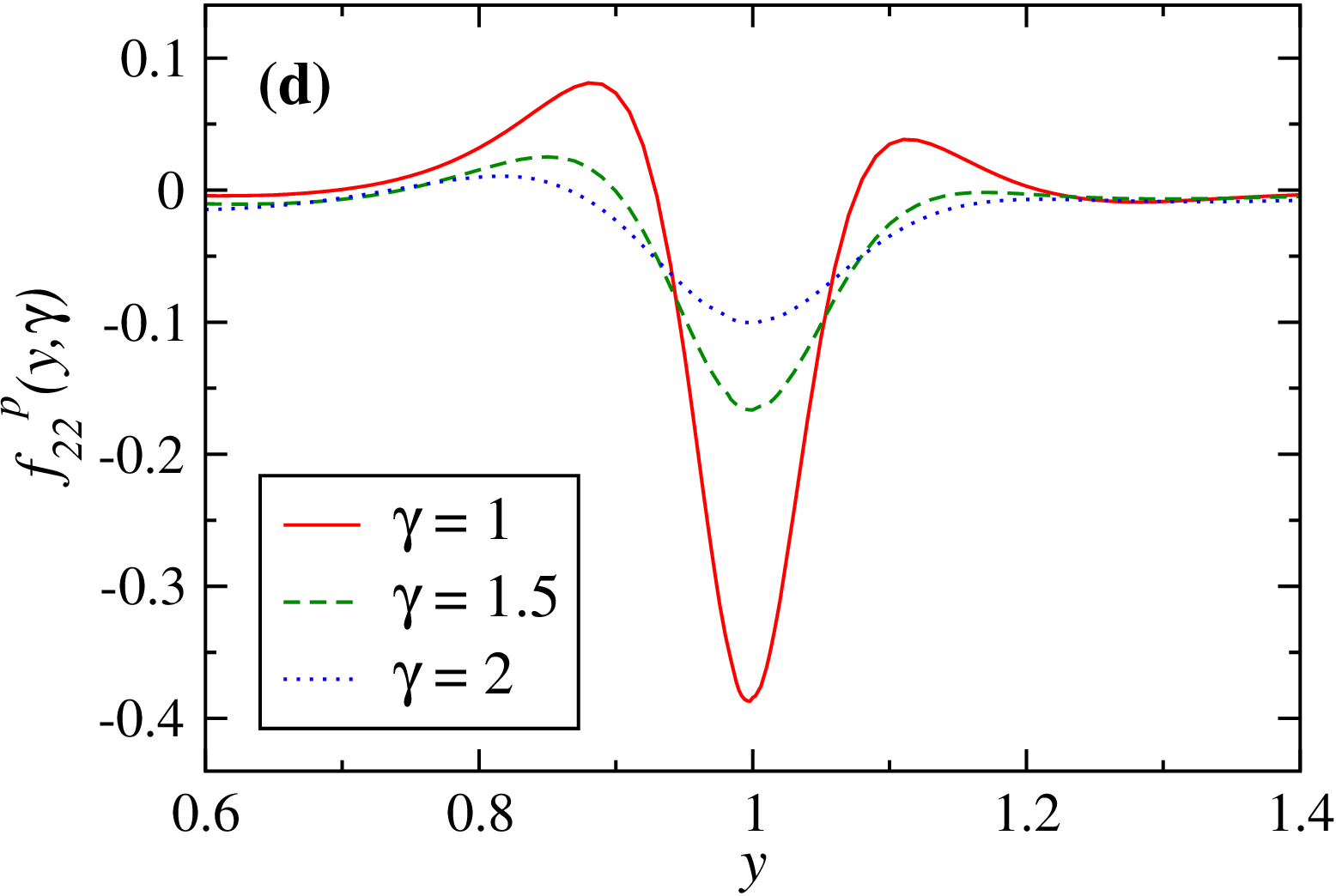,width=9cm}
\caption{(Color online)
	Proton light-cone momentum distribution functions
	$f_{ab}^p(y,\gamma)$ in $^3$He,
	for $\gamma=1$ (Bjorken limit), 1.5 and 2.}
\label{fig:2}
\end{center}
\end{sidewaysfigure}

In Figs.~\ref{fig:1} and \ref{fig:2} we show the neutron and proton
light-cone momentum distribution functions $f_{ab}^\tau(y,\gamma)$ in
$^3$He, respectively, calculated from the SS spectral function \cite{SS},
for several values of $\gamma$ (the results with the KPSV spectral 
function \cite{KPSV} are similar).
For details of the integration over the energy $\varepsilon$ and
momentum $\bm{p}$ in Eq.~(\ref{eq:Dy}) see Ref.~\cite{KP}, Appendix~A.
The results for $\gamma = 1$ correspond to the distributions in the
Bjorken limit.

For the neutron, the diagonal distributions $f_{11}^n$ and $f_{22}^n$
in Figs.~\ref{fig:1}(a) and (d) peak around $y=1$, and drop rapidly
with increasing $|y-1|$.
The finite-$Q^2$ effects render the distributions broader and the
peak smaller with increasing $\gamma$.
The magnitude of the non-diagonal distributions $f_{12}^n$ and
$f_{21}^n$ in Figs.~\ref{fig:1}(b) and (c) is considerably smaller
than that of the diagonal distributions.
In the Bjorken limit the function $f_{12}^n$ in fact vanishes
identically, but becomes finite for $\gamma > 1$, leading to nonzero
contributions from the nucleon $g_2^\tau$ structure function to
$g_1^{^3{\rm He}}$.
The magnitude of $f_{12}^n$ grows with increasing $\gamma$, reaching
at its peak about 6\% of the diagonal $f_{11}^n$ for $\gamma=2$.
The function $f_{21}^n$, on the other hand, is finite for all
$\gamma$, so that the nucleon $g_1^\tau$ structure function
contributes to the nuclear $g_2^{^3{\rm He}}$ even in the Bjorken
limit \cite{ScopHe3}.

For the proton, the $f_{11}^p$ distribution in Fig.~\ref{fig:2}(a)
is negative in the vicinity of $y=1$ and very small in magnitude
compared with the corresponding neutron distribution.
The small proton distribution results from a sizable cancellation
of the (negative) deuteron pole and (positive) $(pn)$ continuum
contributions, as discussed in Sec.~\ref{ssec:wfns}.
The $f_{22}^p$ distribution is qualitatively similar, with a peak
that is negative at $y=1$ and somewhat more pronounced than for
$f_{11}^p$ for $\gamma=1$, but with greater suppression for larger
$\gamma$.
The non-diagonal $f_{12}^p$ distribution for the proton is similar
in magnitude to that for the neutron, vanishing for $\gamma=1$,
while the $f_{21}^p$ distribution remains finite for all $\gamma$.

The strong $\gamma$ dependence of the light-cone momentum distribution
functions will have important consequences for quasi-elastic scattering,
which is given by Eq.~(\ref{eq:convLC}) with the nucleon structure
functions expressed in terms of the elastic nucleon form factors,
\begin{subequations}
\label{eq:QE}
\begin{eqnarray}
g_1^{\tau(\rm el)}(x,Q^2)
&=& {1\over 2}
    { G_M^\tau(Q^2) \left( G_E^\tau(Q^2) + \eta\ G_M^\tau(Q^2) \right)
    \over 1+\eta }\ \delta(x-1)\ ,	\\
g_2^{\tau(\rm el)}(x,Q^2)
&=& {1\over 2}
    { \eta\ G_M^\tau(Q^2) \left( G_E^\tau(Q^2) - G_M^\tau(Q^2) \right)
    \over 1+\eta }\ \delta(x-1)\ ,
\end{eqnarray}
\end{subequations}
where $\eta = Q^2/4M^2$, and $G_E^\tau$ and $G_M^\tau$ are the
Sachs electric and magnetic form factors, respectively.
The presence of the $\delta$-functions in Eqs.~(\ref{eq:QE}) means
that the QE contributions to the structure functions are given by
products of the ($Q^2$-dependent) elastic form factors and the
($x$- and $\gamma$-dependent) light-cone distributions $f_{ab}^\tau$.
For the dominant neutron $f_{11}^n$ distribution, for instance,
a factor of two difference at the QE peak is evident between the
distributions for $\gamma=1$ and $\gamma=2$ (which corresponds
to $Q^2 \sim 1$~GeV$^2$).
A more detailed discussion of QE scattering in the present formalism
will be presented elsewhere \cite{QE}; in the remainder of this paper
we shall focus on inelastic contributions only.

The small overall proton polarization means that the $^3$He nucleus
can, to a good approximation, be used as an effective neutron target,
although for quantitative computations the proton contribution needs
to be accounted for.
If one further neglects Fermi motion and binding so that the $y$
dependence of the distributions $f_{ab}^\tau$ is approximated by
$\delta(y-1)$, and in addition omits the non-diagonal terms $a \neq b$,
then the $^3$He structure functions can be written as simple sums of
proton and neutron structure functions weighted by the effective
nucleon polarizations in Eqs.~(\ref{eq:np-pol}):
\begin{eqnarray}
\label{eq:effpol}
xg_a^{^3{\rm He}}(x,Q^2)
&=& \langle \sigma_z \rangle^p\ xg_a^p(x,Q^2)\
 +\ \langle \sigma_z \rangle^n\ xg_a^n(x,Q^2)\ ,
\end{eqnarray}
for $a = 1,2$.
This simple {\em ansatz} is often used in experimental analyses
to describe spin-dependent $^3$He structure functions.
In the next section we test the accuracy of this {\em ansatz} by
comparing the effects of nuclear smearing using the full results
in Eqs.~(\ref{eq:master}) and (\ref{eq:Dab}) with this and other
approximations, in both the resonance and deep inelastic regions.

\subsection{Results}
\label{ssec:results}

\begin{figure}[t]
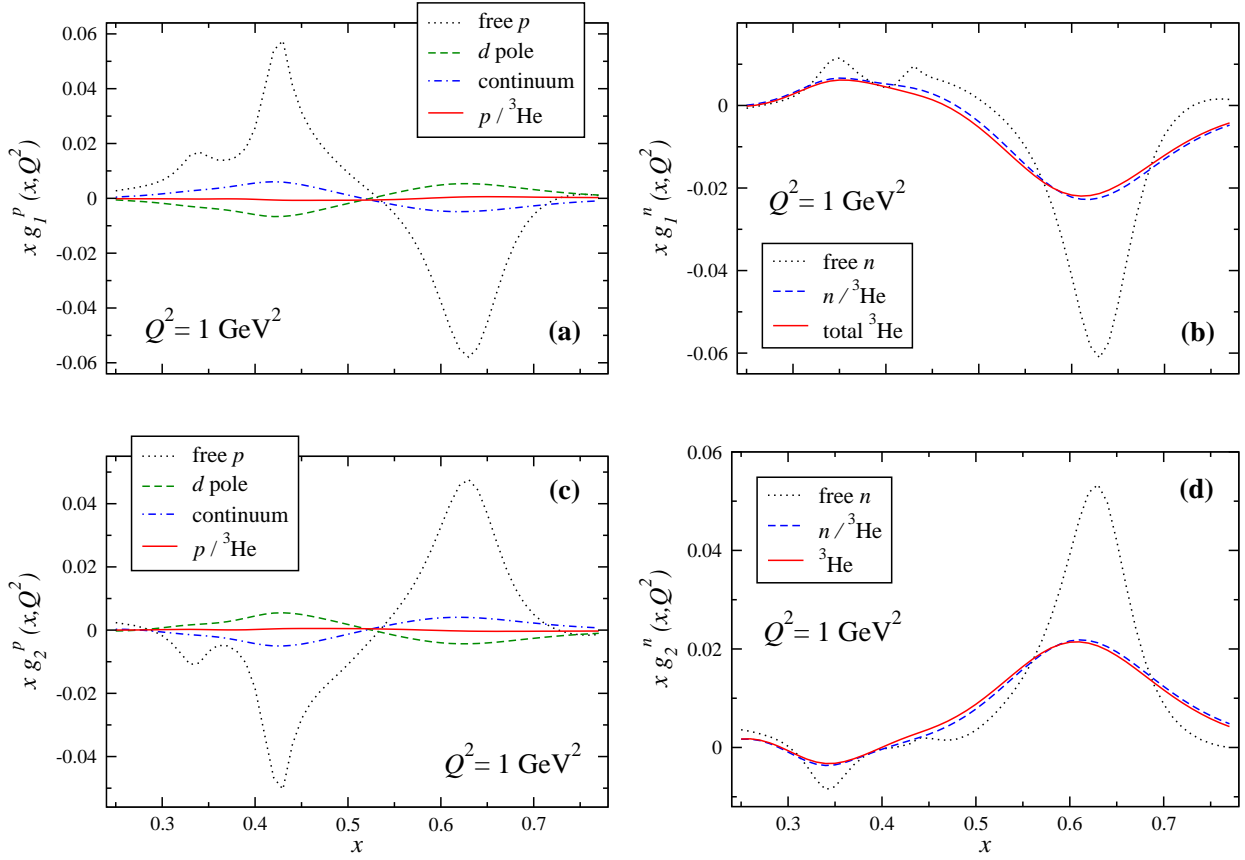

\begin{center}
\epsfig{file=fig3a.eps,width=8cm}\ \ \ %
\epsfig{file=fig3b.eps,width=8cm}\vspace*{0.15cm}
\epsfig{file=fig3c.eps,width=8cm}\ \ \ %
\epsfig{file=fig3d.eps,width=8cm}
\caption{(Color online)
	(a) Proton contribution to $xg_1^{^3{\rm He}}$ (solid),
	including the deuteron pole (dashed) and continuum (dot-dashed)
	contributions, compared with the free proton structure function 
	(dotted).
	(b) Neutron contribution to $xg_1^{^3{\rm He}}$ (dashed),
	compared with the free proton structure function (dotted),
	and the total $xg_1^{^3{\rm He}}$ (solid).
	The MAID parameterization of the free $xg_{1,2}^N$ \cite{N:MAID}
	is used at $Q^2 = 1$~GeV$^2$, with the KPSV $^3$He spectral
	function \cite{Salme}.
	The corresponding proton and neutron contributions to
	$xg_2^{^3{\rm He}}$ are shown in (c) and (d).}
\label{fig:3}
\end{center}
\end{figure}

The calculated $xg_1$ and $xg_2$ structure functions for $^3$He are shown
in Fig.~\ref{fig:3}, using the MAID parameterization \cite{N:MAID} for
the input free proton and neutron structure functions at $Q^2=1$~GeV$^2$,
and the KPSV $^3$He spectral function \cite{Salme}.
The proton contribution to $xg_1^{^3{\rm He}}$ in Fig.~\ref{fig:3}(a)
(solid curve) is considerably smaller in magnitude and of opposite sign
than the free proton structure function (dotted), reflecting the small
negative proton polarization in the $^3$He nucleus, as well as partial
cancellation of the deuteron pole (dashed) and $pn$ continuum
contributions (dot-dashed).

The neutron contribution to $xg_1^{^3{\rm He}}$ in Fig.~\ref{fig:3}(b)
(dashed) shows significant nuclear effects which smear the resonance
peaks in the free neutron $xg_1$ (dotted) and result in a much less
pronounced resonance structure, especially in the $\Delta(1232)$ region
at large $x$.
The similarity of the total $xg_1^{^3{\rm He}}$ (solid) and the neutron
contribution (which arises only from $pp$ continuum states) reflects the
relatively small size of the proton contribution to the $^3$He structure
function.

Similar findings are seen for the proton and neutron contributions to
the $xg_2^{^3{\rm He}}$ structure function in Figs.~\ref{fig:3}(c) and
(d), respectively, which are generally of opposite sign to the $xg_1$
results.

\begin{figure}[t]
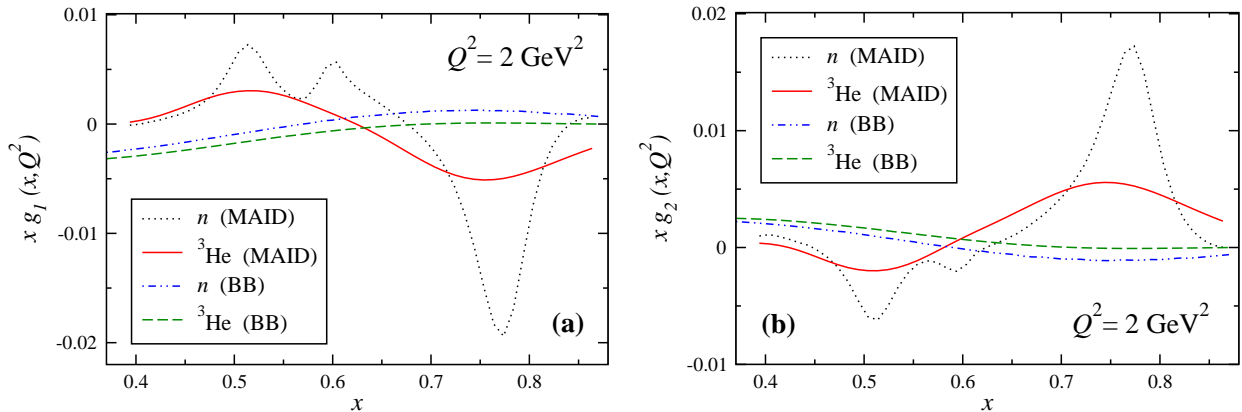
        
\vspace*{1cm}
\begin{center}
\epsfig{file=fig4a.eps,width=8cm}\ \ \ %
\epsfig{file=fig4b.eps,width=8cm}
\caption{(Color online)
	(a) $xg_1$ structure functions of the neutron and $^3$He,
	for the MAID \cite{N:MAID} and BB \cite{N:BB} parameterizations
	at $Q^2 = 2$~GeV$^2$.
	(b) As in (a) but for $xg_2$.
	The $^3$He structure functions are computed with the KPSV
	$^3$He spectral function \cite{Salme}.}
\label{fig:4}
\end{center}
\end{figure}

The effects of the smearing of the neutron structure functions bound in
$^3$He clearly has dramatic consequences for the comparison of free and
bound structure functions.
It considerably complicates, for instance, the extraction of free neutron
structure functions from $^3$He data \cite{Yoni}.
This is in contrast to the effect of smearing in the deep inelastic
region, where the structure functions are smooth, and the differences
between the neutron and $^3$He $xg_{1,2}$ are much less pronounced.
This is illustrated in Fig.~\ref{fig:4}, where the free neutron and
the total $^3$He $xg_1$ and $xg_2$ structure functions are shown in
panels (a) and (b), respectively.
Here the resonance region is parameterized by the MAID model
\cite{N:MAID}, while the deep inelastic curve is given by the leading 
twist fit from Ref.~\cite{N:BB}, both at $Q^2 = 2$~GeV$^2$.

\begin{figure}[ht]
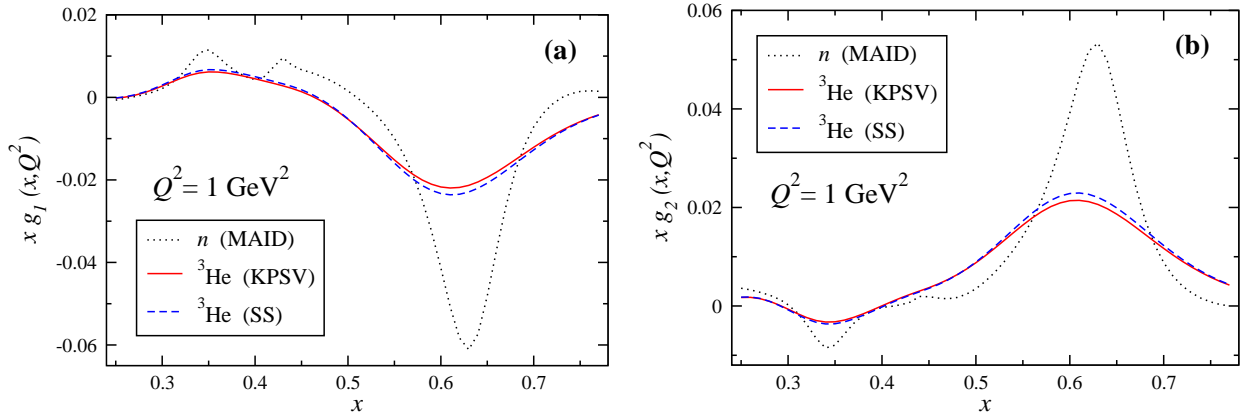

\vspace*{1cm}
\begin{center}
\epsfig{file=fig5a.eps,width=8cm}\ \ \ %
\epsfig{file=fig5b.eps,width=8cm}%
\caption{(Color online)
	Dependence of the (a) $xg_1$ and (b) $xg_2$ structure
	functions of $^3$He on the nuclear spectral function, for the
	KPSV \cite{Salme} (dashed) and SS \cite{SS} (solid) models,
	at $Q^2 = 1$~GeV$^2$.
	The free neutron structure functions from the MAID \cite{N:MAID}
	parameterization are also shown for comparison (dotted).}
\label{fig:5}
\end{center}
\end{figure}

The preceding $^3$He structure functions, in Figs.~\ref{fig:3} and
\ref{fig:4}, have been calculated using the KPSV spectral function
\cite{KPSV,Salme}.
To determine the nuclear spectral function dependence of our 
finite-$Q^2$ smearing, we compare in Figs.~\ref{fig:5}(a) and (b)
the $xg_1^{^3{\rm He}}$ and $xg_2^{^3{\rm He}}$ structure functions,
respectively, for the spectral functions of the KPSV \cite{Salme}
(solid) and SS \cite{SS} (dashed) models.
The free neutron $xg_1^n$ and $xg_2^n$ structure functions from the MAID
parameterization \cite{N:MAID} (dotted) are shown at $Q^2 = 1$~GeV$^2$
for comparison.

Both nuclear models show similarly striking differences between the
free neutron and $^3$He structure functions, demonstrating that the
qualitative features of the smearing are model-independent.
The differences between the two nuclear models are very small, with
about 7\% stronger smearing at the $\Delta(1232)$ resonance peak for the
KPSV spectral function \cite{Salme}, which is due mainly to the slightly
larger neutron polarization than in the SS model \cite{SS}.
The nuclear model dependence in the deep inelastic region, where the
structure functions are smooth, is even smaller.

\begin{figure}[t]
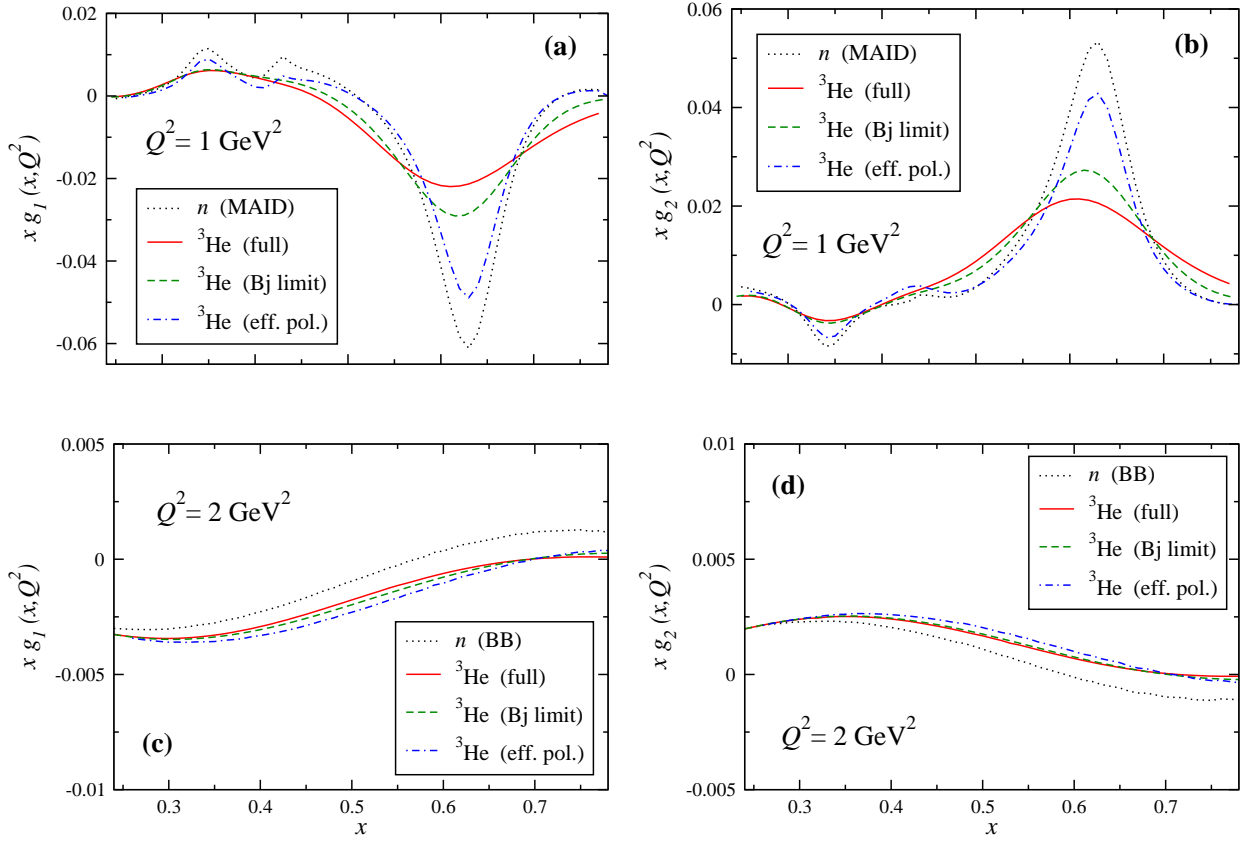

\begin{center}
\epsfig{file=fig6a.eps,width=8cm}\ \ \ %
\epsfig{file=fig6b.eps,width=8cm}\vspace*{0.3cm}
\epsfig{file=fig6c.eps,width=8cm}\ \ \ %
\epsfig{file=fig6d.eps,width=8cm}
\caption{(Color online)
	Comparison of the full calculation of the
	(a) $xg_1^{^3{\rm He}}$ and
	(b) $xg_2^{^3{\rm He}}$
	structure functions (solid), using the KPSV spectral function
	\cite{Salme}, with the Bjorken limit results (dashed),
	and with the effective polarizations approximation (dot-dashed),
	for the MAID \cite{N:MAID} parameterization at $Q^2 = 1$~GeV$^2$.
	Panels (c) \& (d) are as in (a) \& (b), but for the BB
	\cite{N:BB} parameterization at $Q^2 = 2$~GeV$^2$.
	The free neutron structure functions are shown for comparison
	(dotted).}
\label{fig:6}
\end{center}
\end{figure}

To demonstrate the effects of the finite-$Q^2$ smearing, we compare
in Figs.~\ref{fig:6}(a) and (b) the $xg_1^{^3{\rm He}}$ and
$xg_2^{^3{\rm He}}$ structure functions calculated from the full
expressions in Eqs.~(\ref{eq:master}) and (\ref{eq:Dab}) (solid),
using the MAID parameterization \cite{N:MAID} for the free neutron
$xg_{1,2}^n$ at $Q^2 = 1$~GeV$^2$, with the results of taking the
Bjorken limit ($\gamma \to 1$) in the distributions $D_{ab}$ (dashed).
The finite-$Q^2$ results display greater smearing compared with the
$\gamma=1$ case, with some 30\% additional suppression at the
$\Delta(1232)$ resonance peak.

Also shown are the $^3$He structure functions computed using the
effective polarizations in Eq.~(\ref{eq:effpol}).
Not surprisingly, since there is no smearing of the nucleon structure
functions here, the peaks in $xg_{1,2}^{^3{\rm He}}$ for the prominent
resonances are only slightly reduced from those in the neutron,
suggesting that this {\em ansatz} provides a poor approximation to
the full result for the $x$ dependence at finite $Q^2$.
A similar observation was also made in Ref.~\cite{ScopHe3}.

This approximation works better at higher $Q^2$ for leading twist
structure functions, where the effects of smearing are not as severe.
This is illustrated in Figs.~\ref{fig:6}(c) and (d) for
$xg_1^{^3{\rm He}}$ and $xg_2^{^3{\rm He}}$ computed with the
BB leading twist parameterization \cite{N:BB} at $Q^2 = 2$~GeV$^2$.
In this case the effective polarization {\em ansatz} overestimates
the nuclear effects in $^3$He at intermediate $x$
($0.3 \lesssim x \lesssim 0.6$) by as much as 30\% compared with
the free neutron $g_{1,2}^n$.
On the other hand, applying the $Q^2$-dependent smearing or that in
the Bjorken limit results in a smaller difference, around $10-20\%$
at $x \sim 0.5$.
With the ever increasing precision of nuclear structure function data
in modern experiments \cite{He3exp}, it is thus highly questionable 
that either effective polarizations or Bjorken limit convolution are
sufficiently reliably methods to account for nuclear corrections
at $Q^2 \sim$ few-GeV$^2$ scales.

\section{Conclusion}

The $^3$He nucleus has for some time been used as an effective
neutron target in spin-dependent deep inelastic scattering
experiments \cite{He3exp}.
This is justified by the small proton polarization in polarized
$^3$He, as our analysis using several realistic spectral functions 
\cite{SS,KPSV,Salme} has confirmed.
While most previous theoretical analyses have focused on nuclear
effects in the DIS region at large values of the final state invariant
mass $W$ and $Q^2$, the new high-precision data in the relatively
unexplored nucleon resonance region at intermediate $Q^2 \sim 1$~GeV$^2$
has motivated a fresh look at nuclear effects which explicitly take
into account kinematical $Q^2$ corrections.
The main purpose of the present work has been to examine in detail
the role of the nuclear corrections in polarized $^3$He at finite
values of $Q^2$, with emphasis on the resonance region.

Using the weak binding approximation, in which the lepton--nucleon
scattering amplitude is expanded to order $\bm{p}^2$ and $\varepsilon$
in the nucleon momentum and energy, we derive expressions for the
nuclear $g_{1,2}^{^3{\rm He}}$ structure functions in terms of the
nucleon $g_{1,2}^\tau$ structure functions, valid at all $Q^2$.
Unlike earlier analyses using convolution formulas with
$Q^2$-independent nucleon light-cone momentum distributions,
we find that inclusion of kinematical $Q^2$ corrections breaks
this simple factorization, giving rise to effective momentum
distribution functions which explicitly depend on the virtual
photon velocity $\gamma = |\bm{q}|/q_0$.

Our results show the smearing effects of the nucleon light-cone
momentum distributions are significantly more dramatic in the nucleon
resonance region than in the deep inelastic region, consistent with
the earlier findings of Ref.~\cite{ScopHe3}, resulting in
much less pronounced resonance structure in $g_{1,2}^{^3{\rm He}}$,
especially in the $\Delta(1232)$ region.
This poses a greater challenge for the extraction of the free
neutron structure functions from nuclear data in the resonance
region than in DIS kinematics, where the differences between the
neutron and $^3$He structure functions are relatively small.

The main effect that we find in this analysis is the broadening of the
peak in the nucleon light-cone momentum distribution functions with
increasing $\gamma$, leading to additional suppression of the nuclear
structure functions around the resonance peaks at finite $Q^2$
relative to the Bjorken limit results.
For $Q^2 = 1$~GeV$^2$, for example, the finite-$Q^2$ results are
some 30\% smaller in magnitude at the $\Delta(1232)$ peak compared
with the $\gamma=1$ case.

Furthermore, the method of effective polarizations, which involves
no smearing at all, results in only $\sim 10-15\%$ suppression of
the $^3$He structure functions compared with the neutron.
This suggests that the effective polarization {\em ansatz} provides
a rather poor approximation to the $g_{1,2}^{^3{\rm He}}$ structure
functions at finite $Q^2$.
With the high precision of polarized $^3$He structure function data in
recent and upcoming experiments \cite{Slifer,Solvignon}, it is important
therefore to account for the correct $Q^2$ dependence when analyzing
data at moderate $Q^2$ and $W$ kinematics.

The formalism presented here provides the necessary framework in
which the finite-$Q^2$ corrections can be quantified within the
impulse approximation.
The analysis can be extended by considering additional effects beyond
the impulse approximation, such as multiple scattering, or final state
interactions, which may be more relevant in quantitative analyses of data
at low $Q^2$ \cite{Slifer,Solvignon}.
Even within the impulse approximation, the possible modification
of the intrinsic structure of the nucleon in the nuclear medium,
which in our framework is represented by the nucleon off-mass-shell
dependence of the bound nucleon structure functions, should be
taken into account.
While currently not very well constrained, such effects can be 
investigated using the methods described for example in
Refs.~\cite{KP,PolEMC,MPT,KMPW,MST,MSTD,KPW,AKL}.

\begin{acknowledgments}

We thank G.~Salm\`e for providing the KPSV $^3$He spectral function
\cite{KPSV,Salme}, and S.~Scopetta for helpful comments.
W.~M. is supported by the DOE contract No. DE-AC05-06OR23177, under
which Jefferson Science Associates, LLC operates Jefferson Lab.
S.~K. is partially supported by the Russian Foundation for Basic
Research, grants No. 06-02-16659 and 06-02-16353.

\end{acknowledgments}

\appendix
\section{Helicity amplitudes and structure functions}
\label{app:hel}

This appendix summarizes the relations between structure functions and
virtual photon helicity amplitudes, which are useful for extracting
specific structure functions from the hadronic tensor.
For completeness, we consider both the spin-dependent $g_1$ and $g_2$
and spin-averaged $F_1$ and $F_2$ structure functions.

Projecting the hadronic tensor $W^{\mu\nu}$ of a target with spin $S$
onto states with definite photon polarizations, the virtual photon 
helicity amplitudes can be written:
\begin{equation}
\label{eq:W:h}
W^{(h,h')}(S) = e^{(h) *}_\mu\ W^{\mu\nu}(S)\ e^{(h')}_\nu\ ,
\end{equation}
where the polarization vector $e_\mu^{(h)}$ describes a virtual
photon with helicity $h$.
In a reference frame in which the momentum transfer is 
$q=(q_0,\bm{0}_\perp,-|\bm{q}|)$, the polarization vectors are
given by
$e^{(\pm 1)}=(0,1,\pm i,0)/\sqrt2$
for right- ($h=+1$) and left- ($h=-1$) polarized photons, and
$e^{(0)}=(q_z, \bm{0}_\perp, q_0)/Q$
for longitudinally polarized photons, where $Q=(Q^2)^{1/2}$.
Evaluating the amplitudes $W^{(h,h')}$ for specific helicities
$h, h'$, one finds:
\begin{subequations}
\begin{align}
\label{eq:W:h:1}
W^{(\pm 1,\pm 1)}
&= F_1 - S_z\left[g_1+(1-\gamma^2)g_2\right]\ ,\\
\label{eq:W:h:0}
W^{(0,0)}
&= F_L\ =\ {\gamma^2 F_2 \over 2x} - F_1\ ,\\
\label{eq:W:h:01}
W^{(0,\pm 1)}
&= -\frac{Q}{\sqrt 2 q_0}(S_x \pm iS_y)(g_1+g_2)\ ,
\end{align}
\end{subequations}
where $\gamma=|\bm{q}|/q_0$.
We also note that for transversely polarized photons the off-diagonal 
terms vanish, $W^{(1,-1)}=W^{(-1,1)}=0$, and the LT-interference terms
for the right- and left-polarized photons are related as
$W^{(\pm 1,0)}=-W^{(0,\pm 1)}$.

One can further define the structure functions
$w_{3/2} \equiv W^{(1,1)}(S_z=+1)$ and
$w_{1/2} \equiv W^{(1,1)}(S_z=-1)$,
which correspond to the projections of the total photon--nucleon spin
in the photon momentum direction equal to 3/2 and 1/2, respectively.
%
Using the inequality $|w_{1/2}-w_{3/2}| \le w_{1/2}+w_{3/2}$, together
with the Schwarz inequality for the off-diagonal helicity amplitude,
$|W^{(0,1)}|^2 \le W^{(0,0)} W^{(1,1)}$, leads then to the following
constraints on the $g_1$ and $g_2$ structure functions:
\begin{align}
\label{eq:ineq:1}
g_1 + (1-\gamma^2)g_2 &\le F_1\ , 
\\
(\gamma^2-1) (g_1+g_2)^2 &\le 2 R F_1^2\ ,
\label{eq:ineq:2}
\end{align}
where $R=F_L/F_1$ is the ratio of longitudinal to transverse 
structure functions for unpolarized scattering.

\section{Off-shell nucleon electromagnetic tensor}
\label{app:Noff}

Here we present a detailed derivation of the most general structure
of the truncated electromagnetic tensor
$\widehat{\mathcal{W}}_{\mu\nu}(p,q)$ for an off-mass-shell nucleon,
representing the forward Compton scattering of a virtual photon
(momentum $q$) from a nucleon (momentum $q$), with the nucleon
legs amputated.
For earlier discussions of the truncated hadronic tensor see
Refs.~\cite{MST,KPW} and \cite{MPT,KMPW} for unpolarized and
polarized structure functions, respectively.
The hadronic tensor for an on-shell nucleon ($p^2 = M^2$) can be
written in terms of $\widehat{\cal W}_{\mu\nu}$ as:
\begin{eqnarray}
\label{eq:W-What}
W_{\mu\nu}(p,q,S)
&=& \frac12 \Tr \left[
    (\dirac{p}+M)
    (1+\gamma_5\dirac{S})\,
    \widehat\mathcal{W}_{\mu\nu}(p,q) \right]\ ,
\end{eqnarray}
where $S$ is the nucleon spin four-vector, orthogonal to the nucleon
four-momentum $p$, $S \cdot p=0$.

The number of independent structure functions (Lorentz--Dirac
structures) which describe $\widehat{\mathcal{W}}_{\mu\nu}$ is
determined by the requirements of the time-reversal ($T$) and parity
($P$) invariance of the electromagnetic interaction, and hermiticity
($H$) of the electromagnetic current operator,
$J_\mu(x) = J_\mu^\dagger(x)$.
These can be summarized as \cite{MPT}:
\begin{subequations}
\label{eq:TPH}
\begin{eqnarray}
\widehat{\mathcal{W}}_{\mu\nu}(p,q) &\stackrel T=&
\left( {\cal T}\ \widehat{\mathcal{W}}^{\mu\nu}(\trans{p},\trans{q})\
       {\cal T}^\dagger
\right)^*\ ,					\label{eq:T}	\\
\widehat{{\cal W}}_{\mu\nu}(p,q) &\stackrel P=&
{\cal P}\ \widehat{{\cal W}}^{\mu\nu}(\trans{p},\trans{q})\
{\cal P}^\dagger\ ,						\\
\widehat{\mathcal{W}}_{\mu\nu}(p,q) &\stackrel H=&
\gamma_0\ \widehat{{\cal W}}_{\nu\mu}^\dagger(p,q)\ \gamma_0\ ,
\end{eqnarray}
\end{subequations}
where ${\trans{p}}^\mu = p_\mu =(p_0, -\bm{p})$,
${\trans{q}}^\mu=(q_0, -\bm{q})$, and ${\cal T}$ and ${\cal P}$
are time-reversal and parity operators, respectively.
In the Dirac representation they are given in terms of the Dirac
matrices as ${\cal T} = -i\gamma_5 {\cal C}$ and ${\cal P} = \gamma_0$,
where ${\cal C} = i\gamma_0\gamma^2$ is the charge conjugation operator.
The asterisk $^*$ in Eq.~(\ref{eq:T}) denotes complex conjugation.

For an on-shell nucleon the requirements (\ref{eq:TPH}), together with
current conservation and the Dirac equation, result in 2 independent 
Lorentz structures for the symmetric electromagnetic tensor,
$W_{\{\mu\nu\}}=(W_{\mu\nu}+W_{\nu\mu})/2$,
and 2 independent structures for the antisymmetric tensor
$W_{[\mu\nu]}=(W_{\mu\nu}-W_{\nu\mu})/(2i)$.
These are parameterized in terms of the usual structure functions as:
\begin{align}
\label{eq:Wsym:onshell}
W_{\{\mu\nu\}}(p,q)
&= -F_1\, \widetilde g_{\mu\nu} 
 + F_2\, \frac{\widetilde p_\mu \widetilde p_\nu}{p\cdot q}\ ,	\\
\label{eq:Wasym:onshell}
W_{[\mu\nu]}(p,q,S)
&= \frac{M}{p\cdot q}\epsilon_{\mu\nu\alpha\beta}\, q^\alpha
\left[
	S^\beta (g_1 + g_2) 
	- p^\beta\frac{S\cdot q}{p\cdot q}\,g_2
\right]\ ,
\end{align}
where $\epsilon_{\mu\nu\alpha\beta}$ is totally antisymmetric tensor with 
$\epsilon_{0123}=1$ and
\begin{subequations}
\begin{align}
\widetilde g_{\mu\nu} &= g_{\mu\nu} - \frac{q_\mu q_\nu}{q^2},
\\
\widetilde p_\mu &= p_\mu - q_\mu \frac{p\cdot q}{q^2}\ .
\end{align}
\end{subequations}

For an off-shell nucleon, the most general Lorentz--Dirac form
of $\widehat{\mathcal{W}}_{\mu\nu}(p,q)$ can be written as:
\begin{eqnarray}
\label{eq:Dir-exp}
\widehat{\mathcal{W}}_{\mu\nu}(p,q)
&=& C_{\mu\nu}^S\,I
 + C_{\mu\nu\alpha}^V \gamma^\alpha
 + C_{\mu\nu}^P i\gamma_5
 + C_{\mu\nu\alpha}^A \gamma_5 \gamma^\alpha
 + C_{\mu\nu\alpha\beta}^T \sigma^{\alpha\beta}\ ,
\end{eqnarray}
where the coefficients are constructed from the nucleon momentum $p$,
the momentum transfer $q$, the metric tensor $g_{\mu\nu}$, and the
antisymmetric tensor $\epsilon_{\mu\nu\alpha\beta}$.
The requirement of hermiticity ensures that the symmetric,
$C_{\{\mu\nu\}}$, and antisymmetric, $C_{[\mu\nu]}$, combinations
are real for all the coefficients in Eq.~(\ref{eq:Dir-exp}).
The discrete symmetries in Eqs.~(\ref{eq:TPH}) and current conservation
impose further requirements on the individual Dirac structures.

$\bullet$
In the Dirac \emph{scalar} sector, Eqs.~(\ref{eq:TPH}) become:
\begin{subequations}
\label{eq:TPH:S}
\begin{eqnarray}
C_{\mu\nu}^{S*}(p,q) &\stackrel T
=& C^{S\mu\nu}(\trans{p},\trans{q})\ ,\\
C_{\mu\nu}^{S}(p,q) &\stackrel P
=& C^{S\mu\nu}(\trans{p},\trans{q})\ .
%
\end{eqnarray}
\end{subequations}
%
%
There are 4 independent $T$-even and $P$-even symmetric Lorentz
structures in this case:
$g_{\mu\nu},\ p_\mu p_\nu,\ q_\mu q_\nu,\
\text{and}\ p_{\{\mu}q_{\nu\}}$,
where we use the notation
$a_{\{\mu}b_{\nu\}} \equiv (a_{\mu}b_{\nu}+a_{\nu}b_{\mu})/2$ and 
$a_{[\mu}b_{\nu]} \equiv (a_{\mu}b_{\nu}-a_{\nu}b_{\mu})/2$.
The current conservation condition, $q_\mu C^{S\mu\nu}=0$, reduces
the number of independent tensors to 2, for which we choose
$\widetilde g_{\mu\nu}$ and $\widetilde p_\mu \widetilde p_\nu$.

For antisymmetric Lorentz structures, the structure $p_{[\mu}q_{\nu]}$
is $T$-odd.
Furthermore, \eqs{eq:TPH:S} suggest that any antisymmetric $T$-even
structure is necessarily $P$-odd.
The only such structure, {\em i.e.} $\epsilon_{\mu\nu} (pq)$,%
\footnote{We define 
	$\epsilon_{\mu\nu\alpha}(b)
	\equiv \epsilon_{\mu\nu\alpha\beta} b^\beta$
	and
	$\epsilon_{\mu\nu}(ab)
	\equiv \epsilon_{\mu\nu\alpha\beta} a^\alpha b^\beta$
	for any four-vectors $a$ and $b$.}
is forbidden in the electromagnetic interaction (although it appears in
the weak-current interaction for the $F_3$ structure function).
Therefore there are no $T$-even and $P$-even antisymmetric Lorentz
structures in the Dirac scalar sector.

$\bullet$
In the Dirac \emph{vector} sector, from Eqs.~(\ref{eq:TPH}) one has:
\begin{subequations}
\label{eq:TPH:V}
\begin{eqnarray}
C_{\mu\nu\alpha}^{V*}(p,q)
	&\stackrel T=& C^{V\mu\nu\alpha}(\trans{p},\trans{q})\ ,\\
C_{\mu\nu\alpha}^{V}(p,q)
	&\stackrel P=& C^{V\mu\nu\alpha}(\trans{p},\trans{q})\ .
\end{eqnarray}
\end{subequations}
%
%
Here there are 10 independent $T$-even and $P$-even symmetric Lorentz
structures:
$g_{\mu\nu}p_\alpha$, $g_{\mu\nu}q_\alpha$,
$p_{\{\mu}g_{\nu\}\alpha}$, $q_{\{\mu}g_{\nu\}\alpha}$,
$p_\mu p_\nu p_\alpha$, $p_\mu p_\nu q_\alpha$,
$q_\mu q_\nu p_\alpha$, $q_\mu q_\nu q_\alpha$,
$p_{\{\mu}q_{\nu\}}p_\alpha$, and $p_{\{\mu}q_{\nu\}}q_\alpha$.
Of these, 5 are ruled out by current conservation,
$q_\mu C^{V\{\mu\nu\}\alpha}=0$, leaving 5 independent $T$-even
and $P$-even symmetric Lorentz structures.
As a convenient basis we take
$\widetilde g_{\mu\nu} p_\alpha$, $\widetilde g_{\mu\nu} q_\alpha$,
$\widetilde p_{\{\mu} g_{\nu\}\alpha}$,
$\widetilde p_\mu \widetilde p_\nu p_\alpha$,
and $\widetilde p_\mu \widetilde p_\nu q_\alpha$.
Similar to the Dirac scalar case, the antisymmetric $T$-even tensors
are necessarily $P$-odd and therefore do not contribute to the
electromagnetic interaction.

$\bullet$
In the Dirac \emph{pseudoscalar} sector, Eqs.~(\ref{eq:TPH}) imply:
\begin{subequations}
\label{eq:TPH:P}
\begin{eqnarray}
C_{\mu\nu}^{P*}(p,q) &\stackrel T
=& -C^{P\mu\nu}(\trans{p},\trans{q})\ ,\\
C_{\mu\nu}^{P}(p,q) &\stackrel P
=& -C^{P\mu\nu}(\trans{p},\trans{q})\ .
%
\end{eqnarray}
\end{subequations}
There is no $T$-even symmetric solution to Eqs.~(\ref{eq:TPH:P}).
Furthermore, one can verify that antisymmetric $T$-even solutions
can only be $P$-odd.
The only such structure is $p_{[\mu}q_{\nu]}$, however, this cannot
be matched with current conservation.
No $T$-even and $P$-even solutions can therefore be found to
Eqs.~(\ref{eq:TPH:P}), so that the Dirac pseudoscalar sector does
not contribute to the expansion in Eq.~(\ref{eq:Dir-exp}).

$\bullet$
In the Dirac \emph{axial vector} sector, one has:
\begin{subequations}
\label{eq:TPH:A}
\begin{eqnarray}
C_{\mu\nu\alpha}^{A*}(p,q)
	&\stackrel T=& C^{A\,\mu\nu\alpha}(\trans{p},\trans{q})\ ,\\
C_{\mu\nu\alpha}^{A}(p,q)
	&\stackrel P=& -C^{A\,\mu\nu\alpha}(\trans{p},\trans{q})\ .
\end{eqnarray}
\end{subequations}
One immediately observes from Eqs.~(\ref{eq:TPH:A}) that
$C_{\{\mu\nu\}\alpha}^A=0$.
There are 3 independent current conserving structures in the
antisymmetric tensor $C^A_{[\mu\nu]\alpha}$:
$\epsilon_{\mu\nu\alpha}(q)$, $\epsilon_{\mu\nu}(pq)p_\alpha$, and
$\epsilon_{\mu\nu}(pq)q_\alpha$.

$\bullet$
Finally, in the Dirac \emph{tensor} sector, the transformations are:
\begin{subequations}
\label{eq:TPH:T}
\begin{eqnarray}
C_{\mu\nu\alpha\beta}^{T*}(p,q) &\stackrel T=&
	-C^{T\,\mu\nu\alpha\beta}(\trans{p},\trans{q})\ ,\\
C_{\mu\nu\alpha\beta}^{T}(p,q) &\stackrel P=&
	C^{T\,\mu\nu\alpha\beta}(\trans{p},\trans{q})\ ,
\end{eqnarray}
\end{subequations}
from which one concludes that symmetric tensors must vanish,
$C_{\{\mu\nu\}\alpha\beta}^{T}=0$.%
\footnote{There are nontrivial symmetric $T$-even and $P$-odd solutions
	to Eqs.~(\ref{eq:TPH:T}), however, these do not contribute to
	the electromagnetic tensor.}
There are 6 antisymmetric structures $C_{[\mu\nu]\alpha\beta}^{T}$ which
can be constructed from the product of the four-vectors $p$ and $q$ and
the metric tensor $g_{\lambda\sigma}$:
$p_{[\mu}q_{\nu]}p_{[\alpha}q_{\beta]}$,
$p_{[\mu}g_{\nu][\alpha}p_{\beta]}$,
$p_{[\mu}g_{\nu][\alpha}q_{\beta]}$,
$q_{[\mu}g_{\nu][\alpha}p_{\beta]}$,
$q_{[\mu}g_{\nu][\alpha}q_{\beta]}$,
and $g_{\mu[\alpha}g_{\beta]\nu}$.
Furthermore, a number of other possible structures involving bilinear
combinations of the fully antisymmetric tensor can be constructed,
such as
$\epsilon_{\mu\nu\sigma}(q)
 \epsilon_{\alpha\beta}^{\phantom{\alpha\beta}\sigma}(q)$,
$\epsilon_{\mu\nu\sigma}(q)
 \epsilon_{\alpha\beta}^{\phantom{\alpha\beta}\sigma}(p)$,
$\epsilon_{\mu\nu}(pq)
 \epsilon_{\alpha\beta}(pq)$,
{\em etc}.

These structures are not all independent, however.
In particular, one can show that all the Lorentz tensors bilinear
in the antisymmetric tensor $\epsilon_{\mu\nu\alpha\beta}$ can be
rewritten as linear combinations of tensors constructed from 
antisymmetrized products of the vectors $p$ and $q$ and the metric
tensor $g_{\alpha\beta}$.
A direct analysis reveals that there are in fact only 6 independent
combinations obeying Eqs.~(\ref{eq:TPH:T}), which are reduced by
3 additional constraints from the current conservation condition,
$q^\mu C_{[\mu\nu]\alpha\beta}^T=0$.
The 3 remaining basis structures in the Dirac tensor channel are
then chosen to be:
$\epsilon_{\mu\nu\sigma}(q)
 \epsilon_{\alpha\beta}^{\phantom{\alpha\beta}\sigma}(q)$,
$\epsilon_{\mu\nu\sigma}(q)
 \epsilon_{\alpha\beta}^{\phantom{\alpha\beta}\sigma}(p)$,
and
$\epsilon_{\mu\nu}(pq) \epsilon_{\alpha\beta}(pq)$.
In constructing the explicit form of the Dirac expansion in
Eqs.~(\ref{eq:TPH:T}) we also use the identity
$
i\epsilon^{\mu\nu\alpha\beta}\sigma_{\alpha\beta}
= 2\gamma_5\sigma^{\mu\nu}
$.

Collecting the above results, we conclude that the symmetric part
of the truncated nucleon electromagnetic tensor is determined by
the scalar and vector terms in the expansion (\ref{eq:Dir-exp}),
while the antisymmetric part receives contributions from the axial
vector and the tensor terms.
The symmetric tensor $\widehat{\mathcal{W}}_{\{\mu\nu\}}$ generally
involves 7 independent Lorentz--Dirac structures (2 Dirac scalar and 5 
Dirac vector structures), which we write as:
\begin{eqnarray}
\widehat{\mathcal{W}}_{\{\mu\nu\}}(p,q)
&=& -\frac{1}{2M}
    \left( C_S^1\, I + C_V^{1p}{\not{\!p}} + C_V^{1q}{\not{\!q}}
    \right)
    \widetilde g_{\mu\nu}		\nonumber \\
&&+ \frac{1}{2M}
    \left( C_S^2 + C_V^{2p}{\not{\!p}} + C_V^{2q}{\not{\!q}}
    \right)
    \frac{\widetilde p_\mu \widetilde p_\nu}{p\cdot q}
  + \frac{C_V^{\gamma}}{2M}
    \widetilde p_{\{\mu}\widetilde g_{\nu\}\alpha}\gamma^\alpha\ ,
\label{eq:W:sym}
\end{eqnarray}
where the coefficients are real scalar functions of the invariants
$q^2$, $p \cdot q$, and $p^2$, and the normalization factor $1/(2M)$
is introduced to simplify subsequent expressions for the structure
functions.

The antisymmetric tensor $\widehat{\mathcal{W}}_{[\mu\nu]}$ is
constructed similarly from 6 Lorentz--Dirac structures
(3 Dirac axial-vector and 3 Dirac tensor structures),
which can be written as:
\begin{eqnarray}
\widehat{\mathcal{W}}_{[\mu\nu]}(p,q) &=&
\frac{1}{2p\cdot q}\,\epsilon_{\mu\nu\alpha\beta}\,q^\alpha 
\left[
	  C_A^\gamma \gamma_5\gamma^\beta 
	+ p^\beta \gamma_5
	  \left( 
		C_A^p {\not{\!p}} + C_A^q {\not{\!q}} 
	  \right)
\right.\nonumber
\\
&& \left. {}
	+ i p^\beta C_T^{pq} \gamma_5 \sigma^{\rho\lambda}p_\rho q_\lambda
	+ i \gamma_5 \sigma^{\beta\lambda}
	\left(
	C_T^p p_\lambda + C_T^q q_\lambda
	\right)
\right]\ ,
\label{eq:W:asym}
\end{eqnarray}
where again the coefficients are the scalar functions of the
invariants $q^2$, $p \cdot q$, and $p^2$.

Substituting \Eqs{eq:W:sym}{eq:W:asym} into \eq{eq:W-What} we
reproduce \Eqs{eq:Wsym:onshell}{eq:Wasym:onshell} for an on-shell
nucleon, with the structure functions given in terms of the
coefficient functions as:
\begin{subequations}
\label{eq:SF-C}
\begin{align}
F_1 &= C_S^1 + C_V^{1p} M + C_V^{1q} \frac{p\cdot q}{M}\ ,
\\
F_2 &= C_S^2 + C_V^{2p} M
     + \left(C_V^{2q} + C_V^\gamma\right) \frac{p\cdot q}{M}\ ,
\\
g_1 &= -C_A^\gamma + C_T^p - C_A^q p\cdot q - C_T^{pq}Mp\cdot q\ ,
\\
g_2 &= \left(C_A^q M+C_T^q +C_T^{pq}M^2\right)\frac{p\cdot q}{M}\ .
\end{align}
\end{subequations}
Note that the term proportional to $\gamma_5\dirac p$ in \eq{eq:W:asym} 
gives a vanishing contribution to the spin structure functions $g_{1,2}$ 
because of the condition $p \cdot s=0$, so that only five of the
possible six structures in \eq{eq:W:asym} contribute to the physical
nucleon structure functions.

\section{Hadronic tensor in the WBA} 
\label{app:WBA}

In this appendix we discuss the nucleon hadronic tensor in the weak
binding approximation and derive the reduction of the four-dimensional
spinor trace to the two-dimensional trace in \eq{eq:Tr:NR}.
To this end we consider the traces $\Tr [{\cal O A}]$ with the
basis Dirac operators from \Eqs{eq:W:sym}{eq:W:asym}.
Using the notation $p=(M+\ceps,\bm p)$ for the nucleon four-momentum
$p$ and the relation between the four-dimensional and two-dimensional
spinors in \eq{eq:N-psi}, the traces can be written as:
\begin{equation}
\label{eq:NRtr}
\frac{1}{2M_A}
\Tr\left[ {\cal O A}(p,S) \right]
=
\frac{1}{p_0}
\tr\left[ {\cal O}^{\text{WBA}}{\cal P}(\ceps,\bm p,\bm S) \right]\ ,
\end{equation}
where the relativistic (${\cal A}$) and nonrelativistic (${\cal P}$)
spectral functions are given by \Eqs{eq:A}{eq:spfn}, respectively.
The results for the operators ${\cal O}^{\text{WBA}}$ in the weak
binding approximation are listed in Table~\ref{tab:NRtr}.
These results are derived by systematically expanding in $1/M$ and
keeping terms to order $\bm p^2/M^2 \sim \ceps/M$.
To this order the nucleon 3-momentum and off-shell mass are related
by $1 - \bm p^2/(2M^2) = \sqpsq / p_0$.

\begin{table}[h]
\begin{center}
\begin{tabular}{|c|c|}
\hline
${\cal O}$ & $\quad {\cal O}^{\text{WBA}}$ \\
\hline 
$I$ 			& $\quad \sqpsq$ \\
$\gamma^\alpha$ 	& $\quad p^\alpha$ \\
$\gamma_5\gamma^\alpha$ & $-\widehat{\cal S}^\alpha \sqpsq$ \\
$i\gamma_5\sigma^{\alpha\beta} p_\beta$ 
			& $\quad \widehat{\cal S}^\alpha p^2$ \\
$i\gamma_5\sigma^{\alpha\beta} q_\beta$ 
			& $\quad (p\cdot q) \widehat{\cal S}^\alpha
			  - (\widehat{\cal S}\cdot q) p^\alpha$ \\
\hline
\end{tabular}
\caption{Nonrelativistic transformation of the basis Dirac structures.
	The spin operator $\widehat{\cal S}$ is defined by \eq{eq:S:NR}.}
\label{tab:NRtr}
\end{center}
\end{table}

Using \eq{eq:NRtr} together with Table~\ref{tab:NRtr} then leads to
\eq{eq:Tr:NR} for the generic hadronic tensor $\widehat{\cal W}_{\mu\nu}$
discussed in Appendix~\ref{app:Noff}.
The spin-dependent nucleon hadronic tensor $\widehat{w}_{\mu\nu}$
is given by \eq{eq:W:NR} (a similar analysis of the unpolarized case
is given in Ref.~\cite{KP}).
The structure functions for an off-shell nucleon with momentum $p$
are then given by:
\begin{subequations}
\label{eq:SFoff}
\begin{align}
F_1 &= C_S^1\frac{\sqpsq}{M} + C_V^{1p}\frac{p^2}{M}
     + C_V^{1q}\frac{p\cdot q}{M}\ ,
\\
F_2 &= C_S^2\frac{\sqpsq}{M} + C_V^{2p}\frac{p^2}{M} + 
	\left(C_V^{2q}+C_V^\gamma\right)\frac{p\cdot q}{M}\ ,
\\
g_1 &= -C_A^\gamma\frac{\sqpsq}{M} + C_T^p\frac{p^2}{M}
	-\left(C_A^{q}\sqpsq + C_T^{pq}p^2\right)\frac{p\cdot q}{M}\ ,
\\
g_2 &= \left(C_A^{q}\sqpsq + C_T^{pq}p^2 + C_T^q\right)\frac{p\cdot q}{M}\ ,
\end{align}
\end{subequations}
where the coefficients are scalar functions of $p\cdot q,\ Q^2$,
and $p^2$.
One can easily verify that in the limit $p^2 \to M^2$, \eqs{eq:SFoff}
reduce to their correct on-shell limits in \eqs{eq:SF-C}.
Note that the expressions in \eqs{eq:SFoff} are valid in the vicinity
of the nucleon mass shell where nucleon virtuality $p^2-M^2$ is small.
Finally, we observe that although in general the off-shell nucleon
tensor is described by 6 (7) independent structure functions for
spin-dependent (spin-averaged) scattering, it can nevertheless be
characterized by the same number of independent structure functions
as on-shell.


\section{Nuclear spectral function in terms of wave functions} 
\label{app:WF}

Here we discuss the relations between the operator definition of
the nuclear spectral function (\ref{eq:spfn}) and a more traditional
definition in terms of the matrix elements of the wave functions,
as used for instance in Refs.~\cite{KPSV,SS}.
Note that the Fock state of a nucleus $A$ containing nonrelativistic
bound nucleons moving with momentum $\bm P$ can be written as a
product of the nucleon creation operators acting on the vacuum state
convoluted with the nuclear wave function:
\begin{equation}\label{WF:Fock}
|A,\,\bm P \rangle =
\frac{1}{\sqrt{A!}}\int [\ud r]_A \Psi_{\bm P,A}(\{r\}_A)\,
\psi^\dagger(1) \cdots \psi^\dagger(A) |0\rangle\ ,
\end{equation}
where $\psi(i)=\psi_{\sigma_i}^{\tau_i}(\bm r_i)$ is a short-hand
notation for the nucleon field operator at coordinate $\bm r_i$
with polarization $\sigma_i$ and isospin $\tau_i$.
The nuclear wave function $\Psi_{\bm P,A}$ depends on the set of
coordinates, spin, and isospin of the $A$ nucleons, which are
symbolically denoted by $\{r\}_A$, and $[\ud r]_A$ is a symbolic
notation for the integration over coordinates and the sum over
the spin and isospin variables. 
The nuclear wave function also depends on the nuclear momentum as
well as on other quantum numbers, including nuclear spin and isospin,
which are symbolically denoted by $A$.
For a nonrelativistic system the dependence of the wave function on
the center-of-mass momentum and intrinsic variables can be factorized
according to:
\begin{equation}\label{WF:CM}
\Psi_{\bm P,A}(\{r\}_A) = \exp(i\bm P\cdot \bm R_A) \Phi_A(\{\rho\}_A)\ ,
\end{equation}
where
$\bm R_A=\sum_{i=1}^A \bm r_i/A$ is the position of the center-of-mass of 
$A$ particles (we neglect the mass difference for protons and neutrons), 
and the coordinate $\bm\rho_i=\bm r_i - \bm R_A$ describes the position
of the $i$-th particle relative to the nuclear center-of-mass.
In the set of $\bm\rho_i$ only $A-1$ coordinates are independent because 
of the condition $\sum_{i=1}^{A} \bm\rho_i=0$.
The intrinsic wave function $\Phi_A$ is independent of the nuclear 
momentum $\bm P$ and depends on the relative distances between bound
nucleons.
Note also that the integration in \eq{WF:Fock} can be written in terms
of the integration over the center-of-mass position and intrinsic
coordinates as 
$[\ud r]_A
= \ud\bm R_A\ud\bm\rho_1\cdots\ud\bm\rho_A\delta(\sum_{i=1}^{A}\bm\rho_i)$.

We now consider the matrix elements $\psi_f(\bm p,\sigma,\tau)$
in \eq{eq:spfn}.
Using translational invariance we can write
\begin{equation}\label{WF:me1}
\psi_f(\bm p,\sigma,\tau)
= \frac{1}{V}\int\ud\bm r e^{-i\bm p\cdot \bm r}
  \langle(A-1)_f,-\bm p|\psi_\sigma^\tau(\bm r)|A\rangle\ ,
\end{equation}
where $V = \int\ud\bm r$ is normalization volume.
Applying \eq{WF:Fock}, we can then write \eq{WF:me1} as the overlap
integral of the wave functions of the nuclear states.
Using antisymmetry of the wave functions under permutation of bound 
nucleons, we then have
\begin{align}\label{WF:me2}
\psi_f(\bm p,\sigma_1,\tau_1)
= \frac{\sqrt A}{V}\int \ud\bm r_1[\ud r]_{A-1}
  e^{-i\bm p\cdot \bm r_1} \Psi_{-\bm p,(A-1)_f}^*(\{r\}_{A-1})
  \Psi_{\bm 0,A}(\bm r_1,\sigma_1,\tau_1;\{r\}_{A-1})\ ,
\end{align}
which corresponds exactly to the matrix elements in the definition
of the nuclear spectral function in Refs.~\cite{SS,KPSV}.

In the case of three-nucleon system, the Jacobi coordinates
$\bm x=\bm r_3-\bm r_2$ and $\bm y=(\bm r_2+\bm r_3)/2 - \bm r_1$
are often chosen as independent variables for the wave function.
Using \eq{WF:CM} we separate the center-of-mass motion of the residual
two-nucleon state with the center-of-mass coordinate
$\bm R_2=(\bm r_2+\bm r_3)/2$, which leads to:
\begin{align}\label{WF:me3}
\psi_f(\bm p,\sigma_1,\tau_1)
= \sqrt3\int\ud\bm x\ud\bm y\ 
  e^{i\bm p\cdot\bm y}\,
  \Phi_{2f}^*(\bm x;\sigma_2,\sigma_3;\tau_2,\tau_3)
  \Phi_3(\bm x,\bm y;\sigma_1,\sigma_2,\sigma_3;\tau_1,\tau_2,\tau_3)\ ,
\end{align}
where $\Phi_{2f}$ and $\Phi_3$ are the center-of-mass wave functions
of the residual two-nucleon system and the three-body bound state,
respectively.
The sum over the spin and isospin variables $\sigma_2,\sigma_3$ and
$\tau_2,\tau_3$ is implicitly understood.
The spectrum of the residual two-body states is characterized by a
set of quantum numbers $f$, which includes the energy $E_f$, orbital
angular momentum ($L_f$, $L_{zf}$) and spin ($S_f$, $S_{zf}$),
as well as the isospin ($T_f$, $T_{zf}$).

In the case of the deuteron residual state, $\Phi_{2f}$ corresponds
to the deuteron wave function with total angular momentum 1, spin 1
and isospin 0.
For the other channels, the function $\Phi_{2f}$ describes a
continuum two-nucleon state.
The details of calculation of $^3$He and $^3$H bound state wave
function and spectral function can be found in Refs.~\cite{SS,KPSV}
and references therein.



\begin{thebibliography}{99}

\bibitem{EMC}
J.~J.~Aubert {\it et al.}, 
Phys.\ Lett.\  B {\bf 123}, 275 (1983);
Nucl.\ Phys.\  B {\bf 293}, 740 (1987).

\bibitem{EMCrev}
M.~Arneodo,
Phys.\ Rept.\  {\bf 240}, 301 (1994);
%
D.~F.~Geesaman, K.~Saito and A.~W.~Thomas,
Ann.\ Rev.\ Nucl.\ Part.\ Sci.\  {\bf 45}, 337 (1995);
%
P.~R.~Norton,
Rept.\ Prog.\ Phys.\  {\bf 66}, 1253 (2003).

\bibitem{KP}
S.~A.~Kulagin and R.~Petti,
Nucl.\ Phys.\  A {\bf 765}, 126 (2006).

\bibitem{PolEMC}
I.~C.~Clo\"et, W.~Bentz and A.~W.~Thomas,
Phys.\ Rev.\ Lett.\  {\bf 95}, 052302 (2005);
Phys.\ Lett.\  B {\bf 642}, 210 (2006);
%
J.~R.~Smith and G.~A.~Miller,
Phys.\ Rev.\ C {\bf 72}, 022203 (2005).

\bibitem{He3exp}
P.~L.~Anthony {\it et al.}, 
Phys.\ Rev.\  D {\bf 54}, 6620 (1996);
%
K.~Ackerstaff {\it et al.}, 
Phys.\ Lett.\  B {\bf 404}, 383 (1997);
%
M.~Amarian {\it et al.}, 
Phys.\ Rev.\ Lett.\  {\bf 92}, 022301 (2004);
%
X.~Zheng {\it et al.}, 
Phys.\ Rev.\  C {\bf 70}, 065207 (2004);
%
K.~Kramer {\it et al.},
Phys.\ Rev.\ Lett.\  {\bf 95}, 142002 (2005).

\bibitem{Wol}
R.~M.~Woloshyn,
Nucl.\ Phys.\ {\bf A496}, 749 (1989).

\bibitem{Kap}
L.~P.~Kaptari and A.~Yu.~Umnikov,
Phys.\ Lett.\ B {\bf 240}, 203 (1990);
%
L.~P.~Kaptari, K.~Yu.~Kazakov, A.~Yu.~Umnikov and B.~K\"ampfer,
Phys.\ Lett.\ B {\bf 321}, 271 (1994).

\bibitem{Ciofi93}
C.~Ciofi degli Atti, S.~Scopetta, E.~Pace and G.~Salm\`e,
Phys.\ Rev.\ C {\bf 48}, R968 (1993).

\bibitem{SS}
R.-W.~Schulze and P.~U.~Sauer,
Phys.\ Rev.\ C {\bf 48}, 38 (1993).

\bibitem{MPT}
W.~Melnitchouk, G.~Piller and A.~W.~Thomas,
Phys.\ Lett.\ B {\bf 346}, 165 (1995);
%
G.~Piller, W.~Melnitchouk and A.~W.~Thomas,
Phys.\ Rev.\ C {\bf 54}, 894 (1996).

\bibitem{KMPW}
S.~A.~Kulagin, W.~Melnitchouk, G.~Piller and W.~Weise,
Phys.\ Rev.\ C {\bf 52}, 932 (1995).

\bibitem{SSlc}
R.~W.~Schulze and P.~U.~Sauer,
Phys.\ Rev.\  C {\bf 56}, 2293 (1997).

\bibitem{Bissey}
F.~Bissey, A.~W.~Thomas and I.~R.~Afnan,
Phys.\ Rev.\  C {\bf 64}, 024004 (2001).

\bibitem{GDH}
D.~Drechsel, S.~S.~Kamalov, G.~Krein and L.~Tiator,
Phys.\ Rev.\  D {\bf 59}, 094021 (1999);
%
A.~Airapetian {\it et al.}, 
Eur.\ Phys.\ J.\  C {\bf 26}, 527 (2003).

\bibitem{DFINQ}
S.~A.~Kulagin and W.~Melnitchouk,
Phys. Rev. C {\bf 77}, 015210 (2008).

\bibitem{BG}   
E.~D.~Bloom and F.~J.~Gilman,   
Phys.\ Rev.\ Lett.\  {\bf 25}, 1140 (1970).

\bibitem{MEK}  
W.~Melnitchouk, R.~Ent and C.~Keppel,
Phys.\ Rept.\  {\bf 406}, 127 (2005).

\bibitem{ScopHe3}
C.~Ciofi degli Atti and S.~Scopetta,
Phys.\ Lett.\  B {\bf 404}, 223 (1997).

\bibitem{Shad}
G.~Piller and W.~Weise,
Phys.\ Rept.\  {\bf 330}, 1 (2000).

\bibitem{Golak}
J.~Golak, R.~Skibinski, H.~Witala, W.~Glockle, A.~Nogga and H.~Kamada,
Phys.\ Rept.\  {\bf 415}, 89 (2005).

\bibitem{Slifer}
K.~Slifer {\it et al.}, 
Phys.\ Rev.\ Lett.\ {\bf 101}, 022303 (2008).

\bibitem{HLM80}
Y.~Horikawa, F.~Lenz and N.~C.~Mukhopadhyay,
Phys.\ Rev.\  C {\bf 22}, 1680 (1980).

\bibitem{Friar}
J.~L.~Friar \emph{et al.},
Phys.\ Rev.\ C {\bf 42}, 2310 (1990).

\bibitem{KPSV}
A.~Kievsky, E.~Pace, G.~Salm\`e and M.~Viviani,
Phys.\ Rev.\  C {\bf 56}, 64 (1997).

\bibitem{Salme}
G.~Salm\`e,
private communication.

\bibitem{KVR}
A.~Kievsky, M.~Viviani and S.~Rosati,
Nucl.\ Phys.\ {\bf A551}, 241 (1993).

\bibitem{SSG}
A.~Stadler, P.~U.~Sauer and W.~Gl\"ockle,
Phys.\ Rev.\ D {\bf 44}, 2319 (1991).

\bibitem{QE}
Y.~Kahn, S.~A.~Kulagin and W.~Melnitchouk,
work in progress.

\bibitem{N:MAID}
D.~Drechsel, O.~Hanstein, S.~S.~Kamalov and L.~Tiator,
Nucl.\ Phys.\  A {\bf 645}, 145 (1999).

\bibitem{N:BB}
J.~Bluemlein and H.~Bottcher,
Nucl.\ Phys.\  B {\bf 636}, 225 (2002).

\bibitem{Yoni}
Y.~Kahn, W.~Melnitchouk and S.~A.~Kulagin,
in preparation.

\bibitem{Solvignon}
P.~Solvignon {\it et al.}, 
arXiv:0803.3845 [nucl-ex].

\bibitem{MST}
W.~Melnitchouk, A.~W.~Schreiber and A.~W.~Thomas,
Phys.\ Rev.\  D {\bf 49}, 1183 (1994).

\bibitem{MSTD}
W.~Melnitchouk, A.~W.~Schreiber and A.~W.~Thomas,
Phys.\ Lett.\  B {\bf 335}, 11 (1994);
%
W.~Melnitchouk and A.~W.~Thomas,
Phys.\ Lett.\  B {\bf 377}, 11 (1996).

\bibitem{KPW}
S.~A.~Kulagin, G.~Piller and W.~Weise,
Phys.\ Rev.\  C {\bf 50}, 1154 (1994).

\bibitem{AKL}
S.~I.~Alekhin, S.~A.~Kulagin and S.~Liuti,
Phys.\ Rev.\  D {\bf 69}, 114009 (2004);

\end{thebibliography}
\end{document}